\documentclass[11pt,twocolumn]{article}

\usepackage{amsmath,amsfonts,amssymb,bbm,enumitem,url,color,graphicx,amsthm,rotating,authblk,soul,siunitx,tabulary,booktabs}
\usepackage[font=footnotesize,labelfont=bf]{caption}
\usepackage[a4paper]{geometry}
\usepackage[round]{natbib}

\begin{document}

\twocolumn[	\begin{@twocolumnfalse}
		\begin{center}
			{\bf \Large Tropical transition of Hurricane Chris (2012)}\\
			{\bf \Large over the North Atlantic Ocean:}\\
			{\bf \Large A multi-scale investigation of predictability}
			
			\bigskip
			\medskip
			
			{\bf Michael Maier-Gerber\textsuperscript{*1}, Michael Riemer\textsuperscript{2}, Andreas H.~Fink\textsuperscript{1},\\ Peter Knippertz\textsuperscript{1}, Enrico Di Muzio\textsuperscript{1,2}, Ron McTaggart-Cowan\textsuperscript{3}}\\

			\bigskip
			{\tiny\textsuperscript{1} Institute of Meteorology and Climate Research, Karlsruhe Institute of Technology, Karlsruhe, Germany}\\
			{\tiny\textsuperscript{2} Institute for Atmospheric Physics, Johannes Gutenberg University, Mainz, Germany}\\
			{\tiny\textsuperscript{3} Numerical Weather Prediction Research Section, Environment and Climate Change Canada, Dorval, Quebec, Canada}

			\bigskip
			\footnotesize\textsuperscript{*}michael.maier-gerber@kit.edu
			
			\bigskip
			
			\textbf{This work has been submitted to Monthly Weather Review.\\Copyright in this work may be transferred without further notice.}
			
			\bigskip

		\end{center}

\abstract{Tropical cyclones that evolve from a non-tropical origin may pose a special challenge for predictions, as they often emerge at the end of a multi-scale cascade of atmospheric processes. Climatological studies have shown that the 'tropical transition' (TT) pathway plays a prominent role in cyclogenesis, in particular over the North Atlantic Ocean. Here we use operational European Centre for Medium-Range Weather Forecasts ensemble predictions to investigate the TT of North Atlantic Hurricane Chris (2012), whose formation was preceded by the merger of two potential vorticity (PV) maxima, eventually resulting in the storm-inducing PV streamer. The principal goal is to elucidate the dynamic and thermodynamic processes governing the predictability of cyclogenesis and subsequent TT. Dynamic time warping is applied to identify ensemble tracks that are similar to the analysis track. This technique permits small temporal and spatial shifts in the development.\\
The formation of the pre-Chris cyclone is predicted by those members that also predict the merging of the two PV maxima. The position of the storm relative to the PV streamer determines whether the pre-Chris cyclone follows the TT pathway. The transitioning storms are located inside a favorable region of high equivalent potential temperatures that result from a warm seclusion underneath the cyclonic roll-up of the PV streamer. A systematic investigation of consecutive ensemble forecasts indicates that forecast improvements are linked to specific events, such as the PV merging. The present case exemplifies how a novel combination of Eulerian and Lagrangian ensemble forecast analysis tool allows to infer physical causes of abrupt changes in predictability.\\	
	\bigskip}

\end{@twocolumnfalse}]

\section{Introduction}
\label{sec:Intro}
''Tropical transition'' (TT) describes the phenomenon when a tropical cyclone (TC) emerges from an extratropical cyclone \citep{Davis2003, Davis2004}. During TT, the extratropical cyclone transforms from a cold to a warm core system. A cascade of events commonly precedes the TT: anticyclonic wave breaking \citep[e.g.,][]{Thorncroft1993, Postel1999} causes an upper-level precursor potential vorticity (PV) trough to penetrate into the tropics \citep{Galarneau2015}, which initially induces the development of either an antecedent extratropical \citep{Davis2004} or subtropical cyclone \citep{Evans2009, Gonzalez-Aleman2015, Bentley2016, Bentley2017}. The interplay between the upper-tropospheric PV trough and a low-level baroclinic zone facilitates the organization of convection embedded \citep{Davis2004, Hulme2009} and is characteristic of a TT event, distinguishing it from other baroclinically influenced pathways of TC genesis. The convection associated with the precursor cyclone eventually diminishes the PV gradients above the storm center and hence reduces vertical wind shear \citep{Davis2003, Davis2004}, providing a favorable environment for the cyclone to acquire tropical nature.

TC developments are traditionally classified into 'tropical only' and 'baroclinically influenced' categories \citep[e.g.,][]{Hess1995, Elsner1996}. \cite{McTaggart-Cowan2008, McTaggart-Cowan2013} further suggest a more precise classification of TC development pathways based on an analysis of two metrics, which assess baroclinicity in the lower and upper troposphere, respectively. The two most baroclinically influenced categories are identified to represent ''weak TTs'' and ''strong TTs'', distinguished by the strength of the baroclinicity in the lower troposphere. 

From a climatological perspective, \cite{McTaggart-Cowan2013} reveal that merely \SI{16}{\percent} of all global TCs between 1948 and 2010 resulted from TT, but that the relative importance of the TT development pathway is exceptionally high in the North Atlantic basin (almost \SI{40}{\percent}). Because of the extratropical origin of the precursor PV troughs, North Atlantic TCs that emerge from a TT generally tend to form at higher latitudes \citep{Bentley2016}, and reach weaker intensities on average compared to all TCs \citep{McTaggart-Cowan2008}. However, \cite{Davis2004} point out the challenge of accurately forecasting these events, because they primarily occur in proximity to the eastern seaboard of North America \citep{McTaggart-Cowan2008}. 

In a recent study, \cite{Wang2018} examine reforecasts in terms of their skill to predict tropical cyclogenesis in the North Atlantic, using the pathway classification of \cite{McTaggart-Cowan2013}. The authors conclude that the two TT categories are less predictable than the others, a finding that they attribute to forecast errors of the deep-layer shear and the moisture in the mid-troposphere. They also speculate that the interactions of precursor features and processes required for TT further reduce the overall predictability as those constitute additional probability factors that multiplicatively extend the joint probability for non-TT pathways. Despite this probabilistic way of viewing TT predictability, it still remains unclear which factors of uncertainty cause practical predictability (hereafter predictability always refers to practical predictability and not to the intrinsic predictability of the atmosphere) problems at the different stages of the event. The aim of this study is to fill this gap by identifying (thermo-)dynamic causes for eminently rapid changes in the predictability of a) the occurrence of the pre-Chris cyclone, b) its TT, and c) the structural evolution, in European Centre for Medium-Range Weather Forecasts (ECMWF) ensemble forecasts of Hurricane Chris (2012) throughout the pre-TT portion of the storm's life cycle.

Hurricane Chris was chosen for this multi-scale predictability study because of the complex antecedent PV dynamics and the strong baroclinic environment in the upper and lower levels that facilitated the development of the extratropical precursor cyclone. The storm can be unambiguously classified as a ''strong TT'' case with regard to the climatology of \cite{McTaggart-Cowan2013}, and may therefore serve as a suitable archetype for investigation of TT predictability. At the medium forecast range, the model struggles to develop the pre-Chris cyclone; however, a strong increase in the number of ensemble members predicting cyclone formation two to three days before its actual development in the analysis means that focus can be shifted to the predictability of TT itself from a structural perspective. Cyclone-relative ensemble statistics of consecutive forecasts initialized up to nine days before TT occurs in the analysis will finally allow to link eminently rapid changes in predictability of structural evolution to antecedent (thermo-)dynamic events. Though individual ensemble forecasts have been examined to assess predictability of storm-structural evolution before \citep[e.g.,][]{Gonzalez-Aleman2018}, the present case study is the first to investigate changes in predictability of TT with lead time and thus contributes to a deeper understanding of the associated sources of uncertainty.

Following this introduction, section \ref{sec:Methods} will describe the data used and the methods applied in this study. The synoptic overview in section \ref{sec:Overview} highlights the key atmospheric features that were associated with the TT of Chris, before the results in terms of predictability are presented in section \ref{sec:Results}. The findings and conclusions from this study are discussed in section \ref{sec:Discussion}.

\section{Data and methods}
\label{sec:Methods}
\subsection{Data}
The present case study is based on gridded, six-hourly operational analysis and ensemble forecast data from the European Centre for Medium-Range Weather Forecasts (ECMWF). To assess the evolution of predictability, consecutive ensemble forecasts initialized at 0000 UTC between 10 June 2012 and 19 June 2012 -- equivalent to nine and a half (seven) days prior to TC (the pre-Chris cyclone's) formation -- are systematically investigated. The minimum horizontal grid spacing available for the first 10 days of the ensemble forecasts is \SI{0.25}{\degree}. To allow for comparison, analysis and ensemble forecasts are analyzed using that resolution. PV fields, however, are only examined on the synoptic scale, and thus considered with a coarser resolution of \SI{0.5}{\degree}.

\subsection{Ensemble partitioning: Cyclone vs. no-cyclone}
In a first step, every forecast ensemble considered in this study is split into "cyclone" and "no-cyclone" groups to elucidate dynamic causes limiting predictability of the pre-Chris cyclone's formation. The group memberships are determined based on similarity between forecast tracks and the analysis track using a dynamic time warping technique (see section \ref{ssec:Tracking} for details). A forecast member thus belongs to the "cyclone" ("no-cyclone") group when it predicts (lacks) such a 'similar track'. Group names are italicized hereafter to better distinguish them from text.

\subsection{Cyclone tracking and evaluation of forecast tracks}
\label{ssec:Tracking}
The simple cyclone tracking algorithm described by \cite{Hart2003} is employed here because its performance compared to manual tracking was found to be acceptable. Based on mean sea-level pressure data, this approach successively evaluates \SI{5}{\degree}-squares that partially overlap with their adjacent ones to also consider cyclones at the edges. Three criteria are required to meet for a successful detection of a cyclone center within the square: (1) the minimum central pressure cannot exceed \SI{1020}{hPa}, (2) the square must enclose a 2-hPa gradient, and (3) the center has to be tracked for at least one day. Once all time steps of an analysis or forecast are evaluated in terms of cyclone centers in the domain of interest, further conditions are imposed to connect centers to a physically consistent track. These conditions concern the translation speed and changes in the orientation of the track (see \citealt{Hart2003} for details). More sophisticated tracking algorithms that deal with splits and merges \citep{Neu2013} are not required here because the evolution of the predicted storm is relatively simple in the ensemble.

\begin{figure}
	\centering
	\noindent\includegraphics[width=0.5 \textwidth]{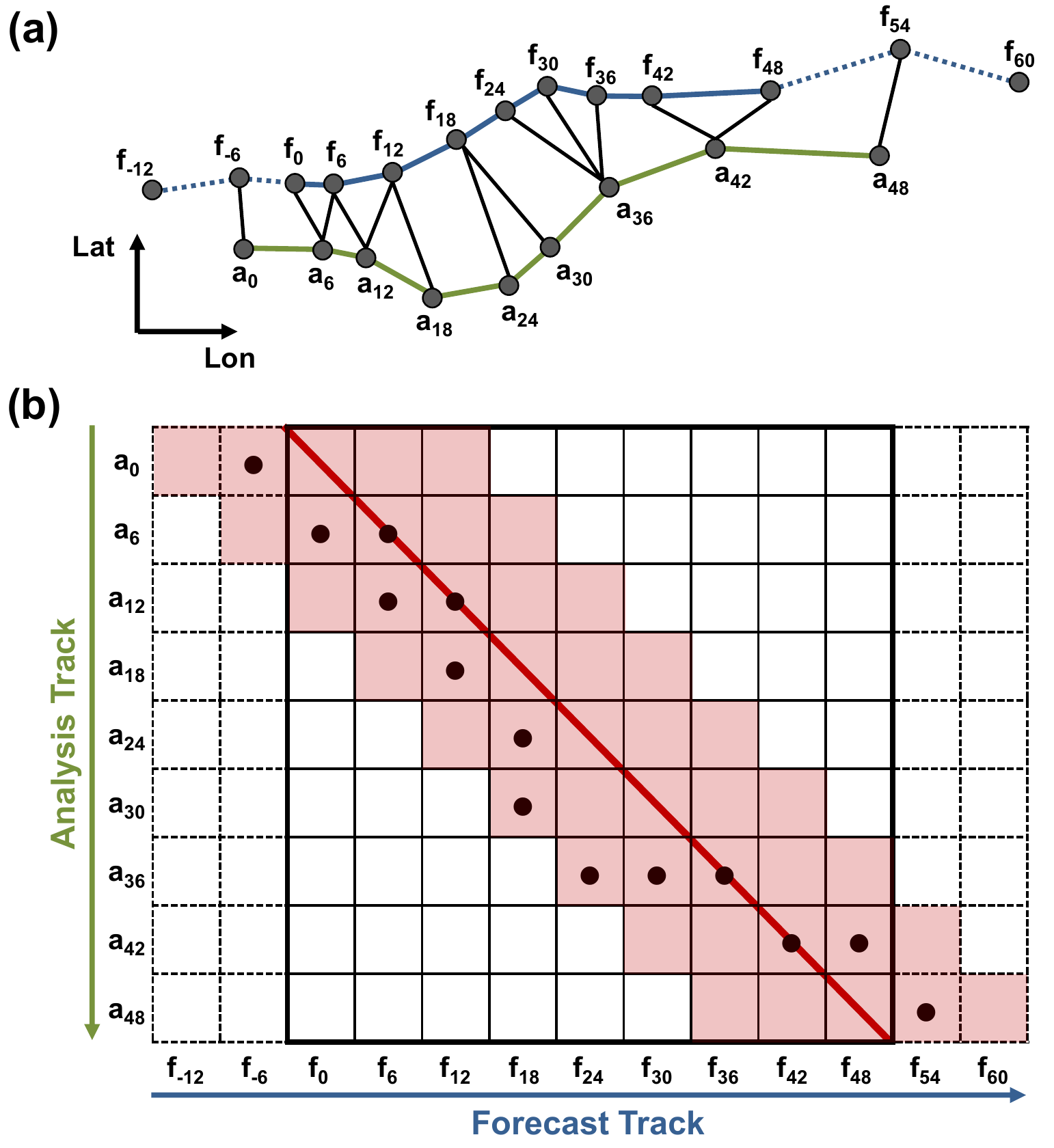}\\
	\caption{(a) Illustrative example of a minimized dynamic time warped path between a forecast track (blue trajectory) and the first \SI{48}{\hour} of the analysis track after the pre-Chris cyclone had formed (green trajectory). The black lines indicate how the total warp path is composed of individual segments that connect the storm positions (gray-filled black circles). (b) Corresponding warping matrix, visualizing how the spatio-temporal similarity is measured. Each matrix element contains a spatial distance between a pair of storm positions from the two storm tracks, with the minimum value identified by a black dot. Because of a maximum allowed temporal shifting of \SI{+-12}{\hour} in the matching, the warping matrix is extended (dashed arrays). The light red area highlights the warping window in which the shortest warp path (connection of the black dots) is allowed to deviate from the non-temporarily shifted matching (red solid line). For more details see section \ref{ssec:Tracking}.}
	\label{f01}
\end{figure}

The application of this tracking technique to the analysis data directly yields a track, which is hereafter referred to as the 'analysis track'. By contrast, an objective approach is required to evaluate whether each forecast member features a track that may be deemed similar to the analysis track. Most studies typically evaluate spatial distances between forecast and analysis tracks, thus making the a priori assumption that forecast errors solely occur in space. However, it may be that a forecast predicts a track accurately in space but slightly shifted in time, for instance due to a lag in the precursor PV dynamics. In the present study, we therefore want to be less restrictive and also allow for small temporal shifts in the forecast tracks, using a dynamic time warping technique. The advantage of this approach is that two tracks of different lengths can be non-linearly matched with respect in time \citep{Sakoe1978,Berndt1994}, thereby taking into account (local) temporal shifts \citep{Chen2004}. Because this case study has a particular focus on the development phase of Chris, the method is applied to calculate the shortest warp path ($d_{DTW}$) between each forecast track and the first \SI{48}{\hour} of the analysis track after the pre-Chris cyclone had formed (not considering cyclone intensity for track matching). In case that a forecast was initialized after the starting time of the analysis track, the first \SI{48}{\hour} after the initialization are considered instead. \cite{Berndt1994} recommend applying a warping window that restricts temporal shifts to ensure reasonable matching. In this case study, forecast tracks are allowed to be locally shifted in time relative to the analysis track by \SI{+-12}{\hour} at most. Figure\,\ref{f01} shows an example of how a forecast track can be aligned with the analysis track, and how this case would be transferred to a time warping matrix, representing the spatial distances of all potential warp path segments between the two tracks. Given the time series of storm positions of the analysis track $A = (a_0, a_6, ..., a_{48})$ and the forecast track $F = (f_{-12}, f_{-6}, f_0, ..., f_{60})$, $d_{DTW}$ can be recursively defined as\\
\begin{equation}
\footnotesize
\begin{split}
d_{DTW}(i,j)= &\| a_{h_i}-f_{h_j}\| \\
&+ Min(d_{DTW}(i-1,j-1),\\
& d_{DTW}(i-1,j),d_{DTW}(i,j-1))
\end{split}
\end{equation}
where the analysis and forecast hours $h_i = 6(i-1)$ and $h_j = 6(j-1)$ are functions of the matrix indices $i=1,2,...,9$ and $j=-1,0,...,11$, respectively. This formulation ensures that the warp path aligns the tracks in a monotonic and continuous manner. In addition to these implicit conditions, some rules are imposed for the boundaries. A candidate forecast track is required to start and end not earlier/later than \SI{+-12}{\hour} compared to the analysis track. For forecast tracks that last longer than the first \SI{48}{\hour} after the formation of the pre-Chris cyclone in the analysis, the later part is not considered in the calculation of $d_{DTW}$. The warping path starts with the distance between the first point of the analysis track ($a_0$) and the closest forecast track point from $(f_{-12}, ..., f_{12})$. Following the recursive definition (1), the calculation of the warp path is continued until the last point of the analysis track $a_{48}$ is aligned with the closest track point of the corresponding forecast subset $(f_{36}, ..., f_{60})$. These boundary conditions guarantee that the analysis track is compared to the most similar part of the forecast track. To allow for comparison with other tracks, the calculated warp path length is then divided by the number of segments to yield the average spatio-temporal discrepancy between forecast and analysis tracks ($\overline{d_{DTW}}$).

All tracks identified in a given ensemble member $m$, are compared to the analysis track following the procedure outlined above. The candidate track with the shortest average warp path is treated as the 'similar track' of ensemble member $m$, provided that $\overline{d_{DTW}}$ does not exceed \SI{700}{\km}. This value was found to be the best choice over a range of thresholds that were tested for all consecutive ensemble forecasts (see Fig.\,1 in the online supplement). Smaller thresholds tend to be too restrictive, higher ones cause an unrealistic inflation of the number of similar tracks.

\subsection{Cyclone group partitioning: Warmer-core vs. colder-core terciles}
\label{ssec:CPSPartitioning}
A further stratification of the \textit{cyclone} group into "warmer-core", "intermediate-core", and "colder-core" terciles allows to investigate whether different tendencies in the ensemble prediction of the cyclone's thermal core-structure can be attributed to distinct (thermo-)dynamic scenarios. Tercile group names are italicized hereafter to better distinguish them from text. In addition, the application of this partitioning strategy to all ensemble forecasts considered (initializations between 10 and 19 June) is also used to link rapid changes in the predictability of structual evolution with lead time to prominent events in the antecedent (thermo-)dynamics. Terciles are technically separated based on the maximum $-V_{T}^U$ values reached between 0000 UTC 19 June and 0000 UTC 20 June (see section 2e for metric definition) to distinguish between storms that barely complete a shallow warm seclusion and TTs that reach the stage of a TC with widespread deep convection. This period of valid times is chosen in preference to a single one to be consistent with the \SI{+-12}{\hour} warping window applied for the dynamic time warping technique (see section 2c). Instead of the tercile-based approach, more sophisticated clustering techniques for trajectories in cyclone phase space (see section 2e) were tested, but had to be rejected, since the resulting clusters either did not separate between distinct physical scenarios, or produced cluster sizes that were too different to proceed with composite analyses.

\subsection{Cyclone phase space}
The cyclone phase space metrics developed by \cite{Hart2003} are calculated along the track to gain insight into the structural evolution of each predicted developing storm and to check whether it transitions into a TC. The result is a set of trajectories in the three-dimensional phase space spanned by the thickness symmetry ($B$), the thermal wind in the lower ($-V_{T}^L$) and in the upper troposphere ($-V_{T}^U$). These three cyclone-relative metrics are derived from geopotential height data within a radius of \SI{500}{\km}. A change in sign of the $-V_{T}^L$ and/or $-V_{T}^U$ metrics is indicative of a change in the thermal structure of the cyclone's core: negative values represent a cold core system and positive values a warm core system. Storm symmetry is assessed by comparing $B$ to the \SI{10}{m} threshold derived from a climatological study by \cite{Hart2003}. Unlike the original definition of the lower (900--600-hPa) and upper layers (600--300-hPa) by \cite{Hart2003}, the alternative separation into 925--700-hPa and 700--400-hPa suggested by \cite{Picornell2014} is used. They argue that these levels are more suitable for higher latitudes because of the lower tropopause height, which is further reduced as the storm interacts with the upper-level precursor trough during the pre-tropical phase. In accordance with \cite{Hart2003}, a 24-h moving average is applied to obtain smoother trajectories, facilitating the identification of phase transitions.

\subsection{Composite approaches, normalized differences, and statistical significance testing}
Applying the partitioning strategies described in section 2b and 2d, two types of composite approaches are calculated to identify differences between the \textit{cyclone} and \textit{no-cyclone} (\textit{warmer-core} and \textit{colder-core}) subsets. While plain earth-relative composites (Eulerian perspective) are primarily used to examine the antecedent dynamics on the synoptic scale prior to the formation of the pre-Chris cyclone, cyclone-relative composites (Lagrangian perspective) provide insight into thermodynamics and convective organization on the mesoscale once the pre-Chris cyclone exists.

Where absolute values matter (e.g. vertical wind shear), composite means are presented for each subset separately. In most cases, however, it is more convenient to directly analyze and discuss composite differences between the separated subsets. Therefore, similar to a forecast sensitivity study from \cite{Torn2015}, normalized differences are computed by subtracting the mean of one subset from the mean of the counterpart subset and subsequently dividing by the ensemble standard deviation to allow for spatial and temporal comparisons. Throughout this paper, statistical significance of composite differences is determined using a bootstrap method with $n=10000$ random draws to resample the unknown underlying probability density functions. For each grid point, it is tested whether the differences are significantly different from zero at the two-sided significance level of \SI{5}{\percent}. To circumvent potential misinterpretations of multiple hypothesis testing \citep{Wilks2016} and to preserve spatial and temporal correlation structures, the resampling is only performed once for every partitioning strategy, i.e. statistical significance testing is based on the same resampling output for every horizontal grid point and forecast time.

\section{Synoptic overview}
\label{sec:Overview}
This section describes the antecedent synoptic-scale dynamics that led to the formation and subsequent TT of Chris, and introduces the key atmospheric features whose predictability will be investigated. In addition, the evolution of the storm's structure during TT is discussed using Hart's cyclone phase space.

\subsection{Formation of the precursor trough}

\begin{figure*}
	\centering
	\noindent\includegraphics[width = 1 \textwidth ]{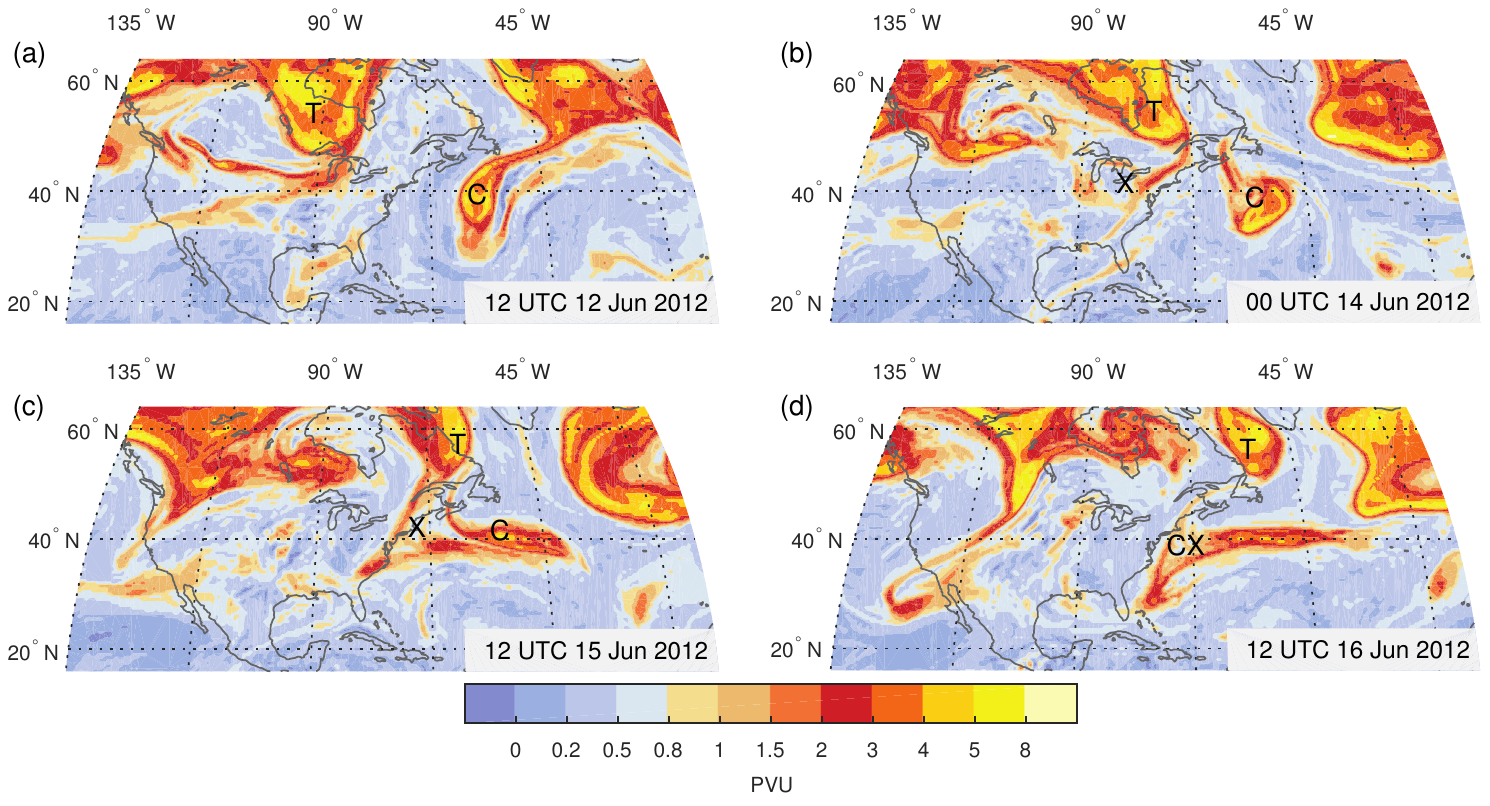}\\
	\caption{Upper-level PV development prior to Chris' formation. 500--250-hPa layer-averaged PV (shaded in PVU, where 1 PVU = \SI{1e-6}{K.m^{2}.kg^{-1}.s^{-1}}) in the ECMWF analysis valid at (a) 1200 UTC 12 June 2012, (b) 0000 UTC 14 June 2012, (c) 1200 UTC 15 June 2012, and (d) 1200 UTC 16 June 2012. The cut-off low, trough, and PV remnants are labeled with C, T, and X, respectively. The merged PV streamer is labeled with CX in (d).}
	\label{f02}
\end{figure*}

The initial trigger for the formation of the precursor trough is anticyclonic wave breaking that took place over the northwest Atlantic Ocean on 12 June 2012 (Fig.\,\ref{f02}a; hereafter all dates are in 2012), which first results in a quasi-stationary, upper-level cut-off low near \SI{55}{\degree}W, \SI{40}{\degree}N (labeled C in Fig.\,\ref{f02}a,b). In the meantime, the upstream trough (labeled T) reaches the east coast of North America, bringing with it unorganized high-PV air (labeled X) to its south, which is the upper-level remnant of a strong Pacific cyclone that hit the west coast of North America on 9 June (not shown). This PV remnants start to interact and merge with the cut-off low (hereafter 'merging'/'merger' always refers to these two PV maxima), forming a zonally oriented PV streamer on 15 June (Fig.\,\ref{f02}c). Because this streamer ultimately acts as the precursor trough for Chris, the 15 June PV merger will be studied in detail from a predictability perspective in section\,\ref{ssec:formation}. During the subsequent development of the pre-Chris cyclone, the western portion of the equatorward penetrating PV merger begins to roll up cyclonically (16 June, label CX in Fig.\,\ref{f02}d).

\subsection{Development of the surface low}

\begin{figure*}[h]
	\centering
	\noindent\includegraphics[width = 1 \textwidth ]{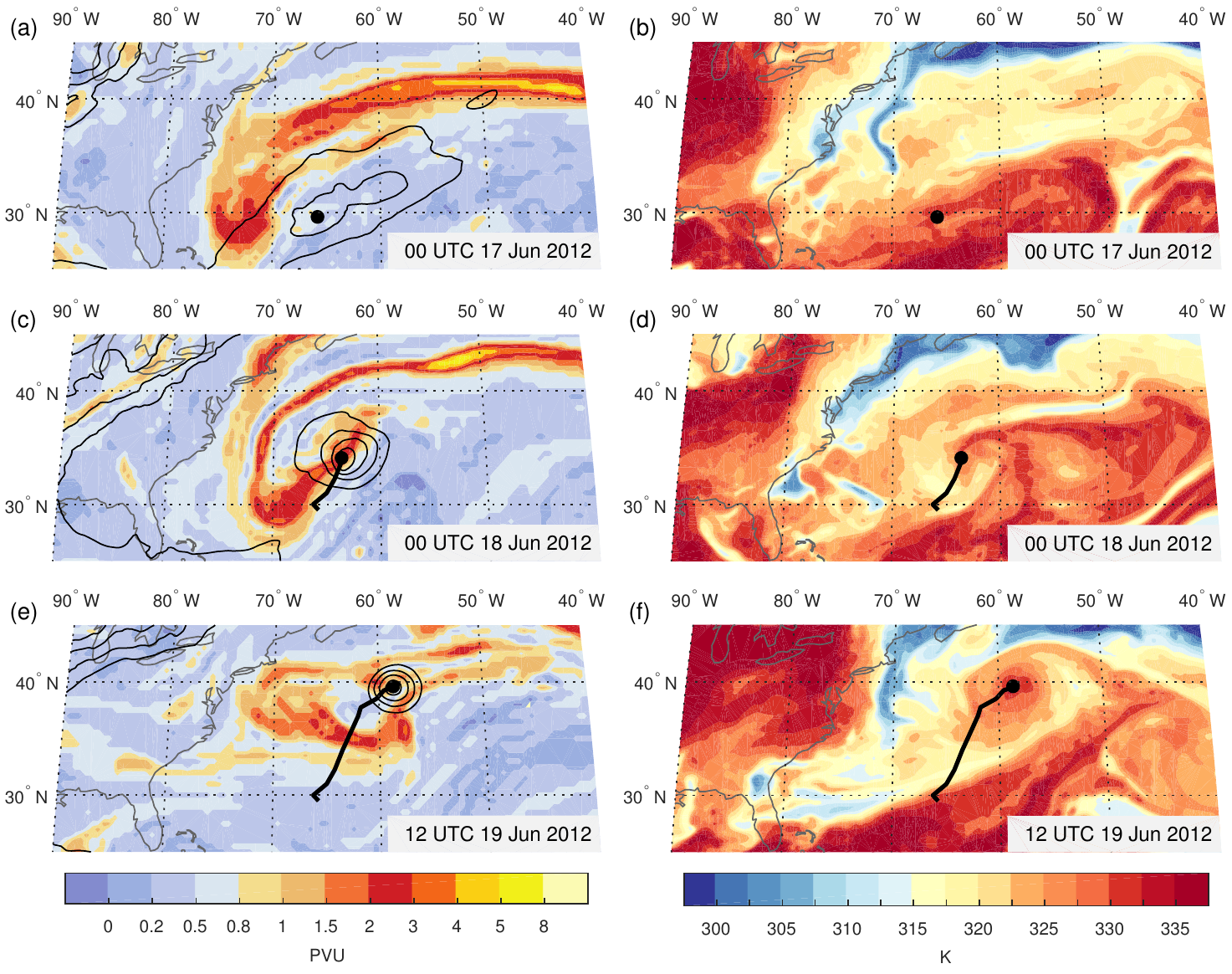}\\
	\caption{Upper-level PV and low-level thermal structure during the pre-tropical phase of Chris. (a),(c),(e) as in Fig.\,\ref{f02}. Additionally, MSLP lower than or equal to 1014 hPa (contours, every 2 hPa) for (a) 0000 UTC 17 June 2012, (c) 0000 UTC 18 June 2012, and (e) 1200 UTC 19 June 2012. (b),(d),(f) $\theta_e$ at 850 hPa (shaded, K) corresponding to the times of (a),(c),(e). Current storm position and past track are denoted by black-filled circles and black lines, respectively.}
	\label{f03}
\end{figure*}

A weak surface low, which would later become Chris, developed during the cyclonic roll-up of the upper-level PV streamer around 0000 UTC 17 June (Fig.\,\ref{f03}a). The center of the surface low is located at the leading edge of the PV streamer, east of its southwestern tip. It will be shown below (section\,\ref{ssec:formation}) that the shape and position of the PV streamer at this time is another key feature for the predictability of Chris' development. The interaction between the surface low and the upstream trough, leading to a further deformation of the PV streamer, is prominent during Chris' development (Fig.\,\ref{f03}a-c). According to the official TC report on Chris from the National Hurricane Center \citep{Stewart2013}, the upper-level PV streamer eventually shears off the cyclone at 1200 UTC on 19 June (Fig.\,\ref{f03}e) and Chris becomes tropical. Hereafter, we therefore consider this time as the end of TT and the beginning of the tropical phase of Chris' life cycle.

In the lower troposphere, the pre-Chris cyclone develops near an air mass boundary, with its center located in the warm, and moist air south of a large equivalent potential temperature ($\theta_e$) gradient (Fig.\,\ref{f03}b). The cyclone is steered northeastward along the leading edge of the upper-level PV streamer as it intensifies (Fig.\,\ref{f03}c). Over this period, a slot of low-$\theta_e$ air wraps cyclonically around the low center (Fig.\,\ref{f03}d), a common feature in subtropical cyclones \citep[e.g.,][]{Hulme2009, Davis2003}. Over the next 36 h, the pre-Chris cyclone detaches from the high-$\theta_e$ reservoir to its south and east and is then embedded in an area of high-$\theta_e$ air surrounded by lower values at radii beyond approximately \SI{350}{\km} (Fig.\,\ref{f03}f).

\subsection{Thermo-structural changes before the TC development}

\begin{figure*}[h]
	\centering
	\noindent\includegraphics[width = 0.9 \textwidth ]{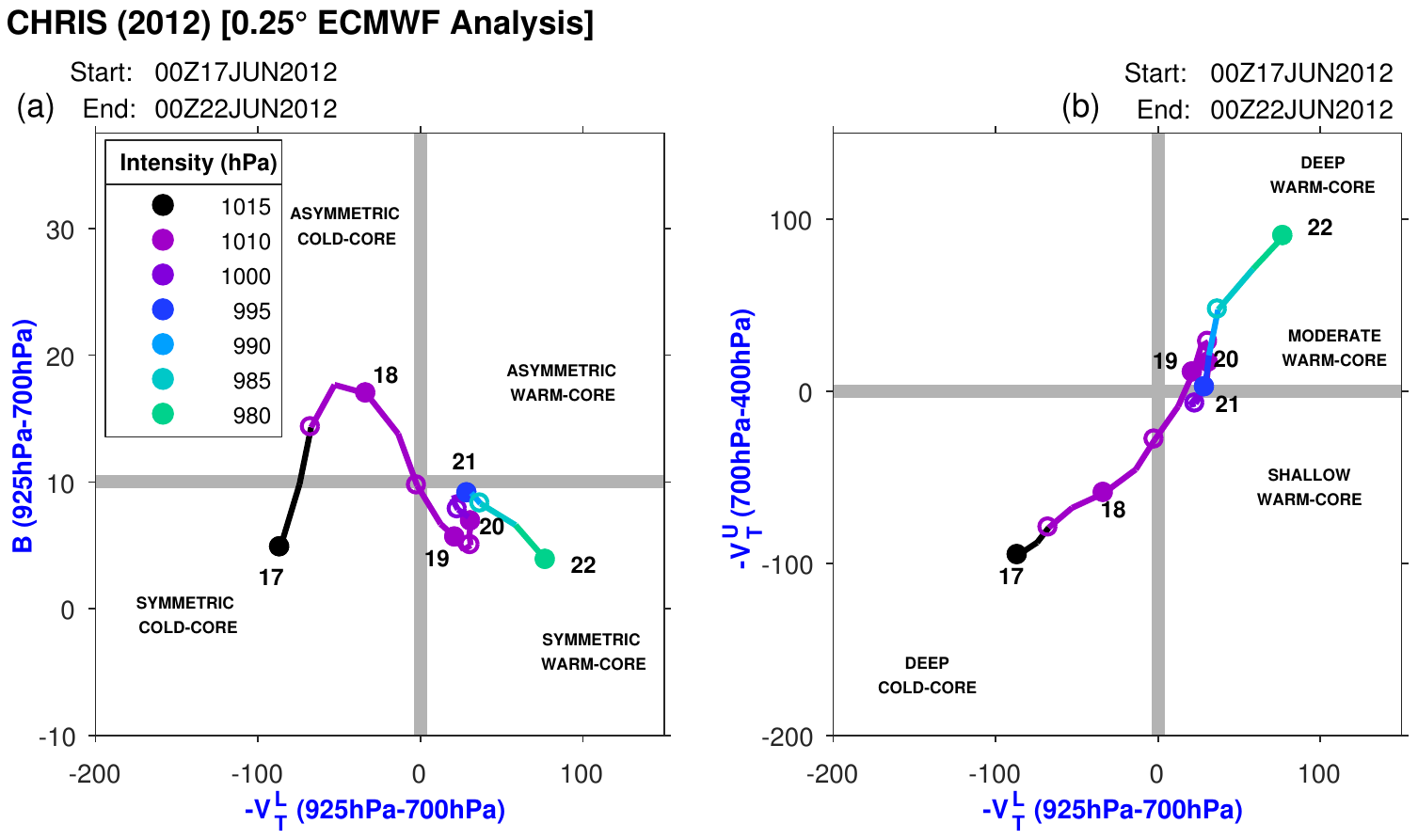}\\
	\caption{Evolution of Hurricane Chris in the CPS. The calculated trajectory for the period from 0000 UTC 17 June 2012 to 0000 UTC 22 June 2012 is projected onto (a) the $B$-$V^L_T$-plane and (b) the $V^U_T$-$V^L_T$-plane. Two thick gray lines in each panel separate between four quadrants, representing different thermo-structural storm stages \citep{Hart2003}. Filled circles denote times at 0000 UTC and open circles at 1200 UTC, respectively.}
	\label{f04}
\end{figure*}

Early in its lifecycle (0000 UTC 17 June), the weak cyclone exhibits a symmetric, cold-core structure (Fig. \ref{f04}a,b). Subsequently, as the surface low starts to interact with the approaching upper-level trough, a wave-like disturbance on the low-level baroclinic zone (Fig.\,\ref{f03}d) causes asymmetric baroclinicity (the $B$ metric) to increase markedly until 0000 UTC 18 June. However, the asymmetric phase ends within \SI{24}{\hour}, with the $B$ metric falling below the $10$-$m$-threshold when the low-level baroclinic zone changes to a warm seclusion with a well-defined dry slot (Fig.\,\ref{f03}f). Over this period of changes in the cyclone-relative symmetry, the core experiences a continuous warming and reaches a symmetric, moderately-deep warm core at 1200 UTC 19 June, consistent with the time at which Chris was declared a tropical storm \citep{Stewart2013}. Regarding thermo-structural changes in the vertical, the CPS diagram reveals a straight transition from a deep cold-core to a warm core of moderate vertical extent (Fig. \ref{f04}b). Both thermal CPS metrics change their signs during the second half of 18 June; $-V_{T}^U$ becomes positive 6 h later than $-V_{T}^L$ as the warm core extends in the vertical over time. Until 0000 UTC 21 June, Chris temporarily interacted with a second trough and became slightly more asymmetric and shallower. Subsequent to this intermediate phase, Chris developed clear tropical characteristics in cyclone phase space, intensified, and became a category 1 hurricane \citep{Simpson1974} at 0600 UTC 21 June \citep{Stewart2013}.

\section{Results}
\label{sec:Results}
\begin{figure}[t!]
	\centering
	\noindent\includegraphics[width = 0.5 \textwidth ]{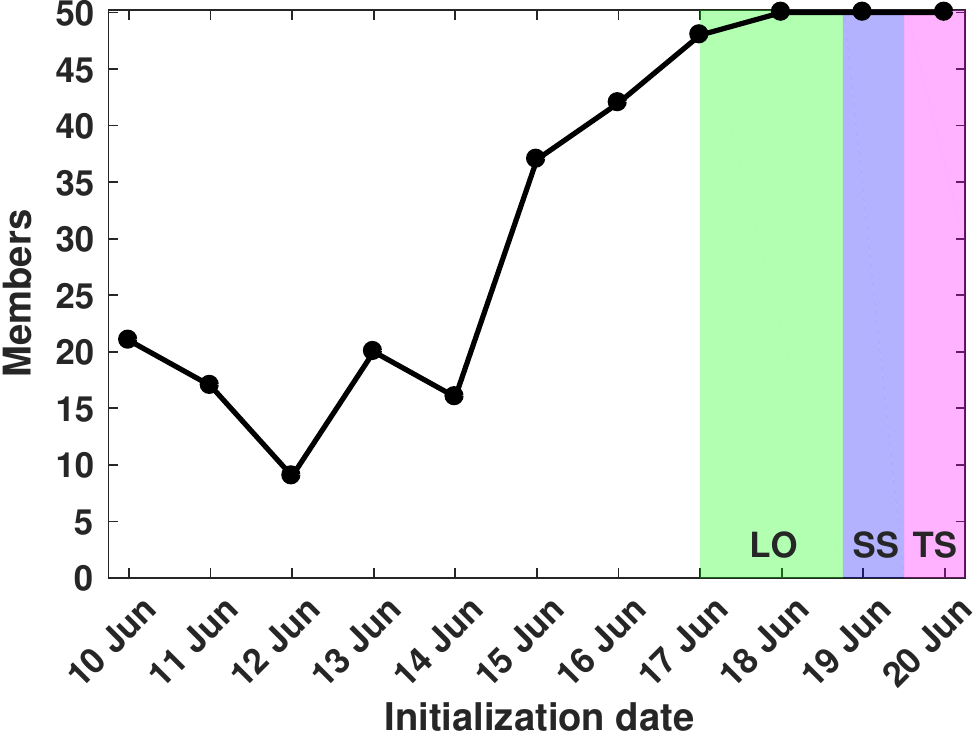}\\
	\caption{Number of ECMWF ensemble members that feature a similar storm track for forecasts initialized at 0000 UTC between 10 June 2012 and 20 June 2012 (for more details on how similarity is measured see section \ref{ssec:Tracking}). The green, blue and red areas highlight the periods when Chris was categorized as a 'low', 'subtropical storm' and 'tropical storm', respectively, by the National Hurricane Center.}
	\label{f05}
\end{figure}

Consecutive ensemble forecasts initialized between 10 and 19 June are examined to address different aspects of predictability. Figure\,\ref{f05} shows how the number of similar tracks evolves as model initialization time gets closer to the tropical stage. An obvious question that arises from this overview is what kept a great portion of the ensemble members from predicting the cyclone before 15 June. As will be shown below, the merging process between the upper-level cut-off low and the PV remnants (features C and X in Fig.\,\ref{f02}), as well as the representation of the overall structure of the resulting precursor trough (CX in Fig.\,\ref{f02}) are major limitations at different lead times. Once the majority of the members predict a similar track, the main focus turns to the TT aspect to explore why some of the ensemble members acquire a more tropical structure, while the others remain less tropical. The section ends with a systematic investigation of predictability of Chris' structural evolution.

\subsection{Predictability of storm formation}
\label{ssec:formation}

\begin{figure*}[h]
	\centering
	\noindent\includegraphics[width = 1 \textwidth ]{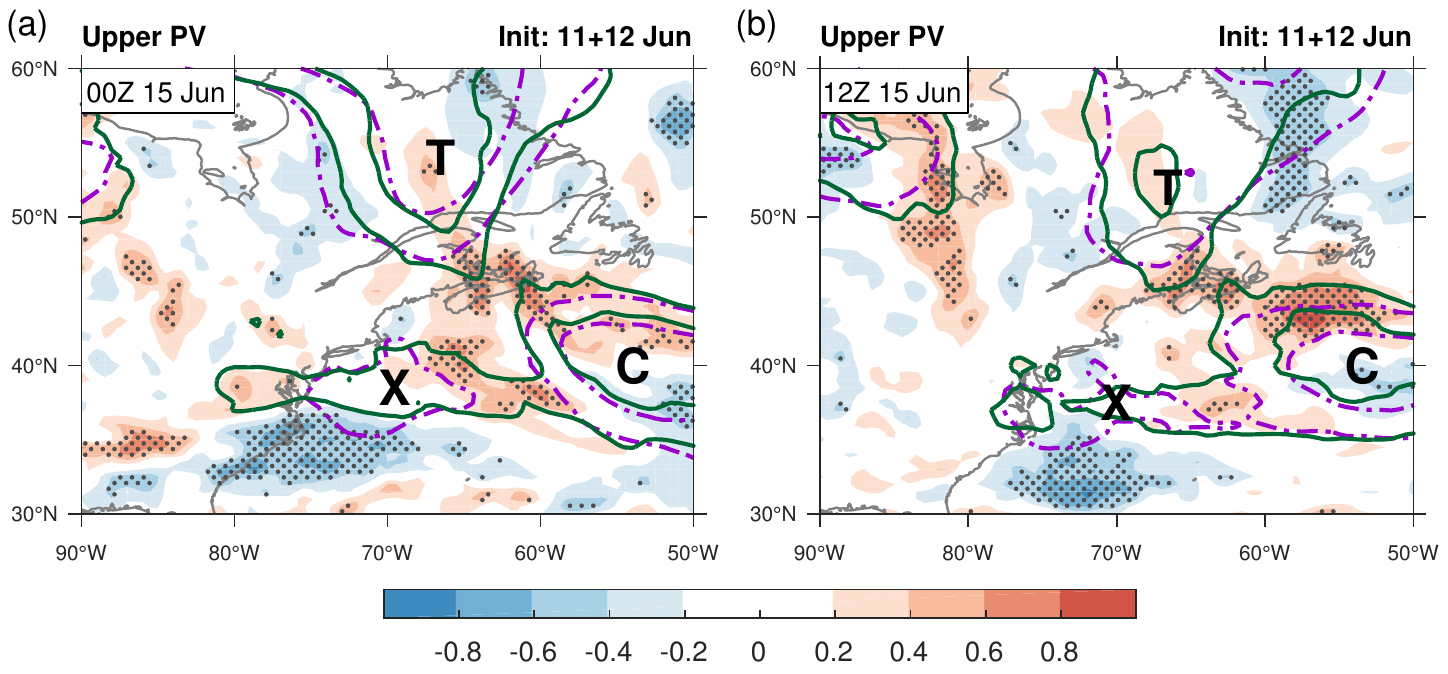}\\
	\caption{Normalized difference in 500--250-hPa layer-averaged PV (shaded, units: standardized anomaly) between the \textit{cyclone} and \textit{no-cyclone} groups of the combined ensemble forecasts from 0000 UTC 11 and 12 June 2012 valid at (a) 0000 UTC 15 June 2012 and (b) 1200 UTC 15 June 2012. Differences significant at the 95\% confidence level are indicated by gray stippling. The solid green \textit{cyclone} group-averaged and dotdashed purple \textit{no-cyclone} group-averaged PV contours (at 1 and 2 PVU) serve as reference for the trough, cut-off low, and PV remnants, which are labeled with T, C, and X, respectively.}
	\label{f06}
\end{figure*}

Figure\ 6 displays normalized differences in 500--250-hPa layer-averaged PV of the combined ensemble forecasts from 0000 UTC 11 and 12 June between the \textit{cyclone} ($\overline{d_{DTW}} \leq$ \SI{700}{\km}) and \textit{no-cyclone} ($\overline{d_{DTW}} >$ \SI{700}{\km}) groups valid at two times on the merging day (15 June). Combining the two initialization times is reasonable, as both predict similar PV structures and areas of anomalies relative to the ensemble mean for the \textit{cyclone} and \textit{no-cyclone} groups, respectively. For forecasts initialized before 11 June, significant differences associated with the merging process are sparse and less coherent because of large spread in the ensemble. After the 12 June initialization, the two groups only exhibit minor differences as the initialization date gets closer to the merging day (not shown). This appears to be because all members share a similar depiction of the merging process at shorter lead times.

At the beginning of the merging day (15 June), the cut-off low (C), the trough (T) and the PV remnants (X) are in closer proximity in the \textit{cyclone} group, whereas they are clearly separated in the \textit{no-cyclone} group (Fig.\,\ref{f06}a). In contrast to the isolated configuration in the \textit{no-cyclone} composite, the cut-off low in the \textit{cyclone} composite is stronger interacting with the trough and has already merged with the PV remnants to the west, as can be seen from the 1-PV unit (PVU, where 1 PVU = \SI{1e-6}{K.m^{2}.kg^{-1}.s^{-1}}) composite isopleths and the significant positive differences between the three PV features. Concerning the strength and location of the interacting PV features, a dipole in the group differences associated with the cut-off (C) reveals that the cut-off in the \textit{cyclone} group is located north of the \textit{no-cyclone} group. In addition, the \textit{cyclone} group predicts positive differences in the center of the trough (T) encompassed by negative values to the east and west. This characterizes a narrower, but more intense trough, reaching slightly farther south.

Over the course of the merging day (15 June), the \textit{cyclone} group shows the cut-off merged with the eastern part of the PV remnants; in contrast they are still separated in the \textit{no-cyclone} group (Fig.\,\ref{f06}b). The cut-off in the \textit{cyclone} composite remains displaced to the north relative to the \textit{no-cyclone} composite and the trough remains sharper. The pronounced PV maximum south of \SI{55}{\degree}N and the large significant negative area north of Newfoundland indicate a more negatively-tilted trough compared to the \textit{no-cyclone} group, favorable for a cyclonic PV roll-up \citep{Shapiro1999}.

\begin{figure}[h]
	\centering
	\noindent\includegraphics[width = 0.45 \textwidth ]{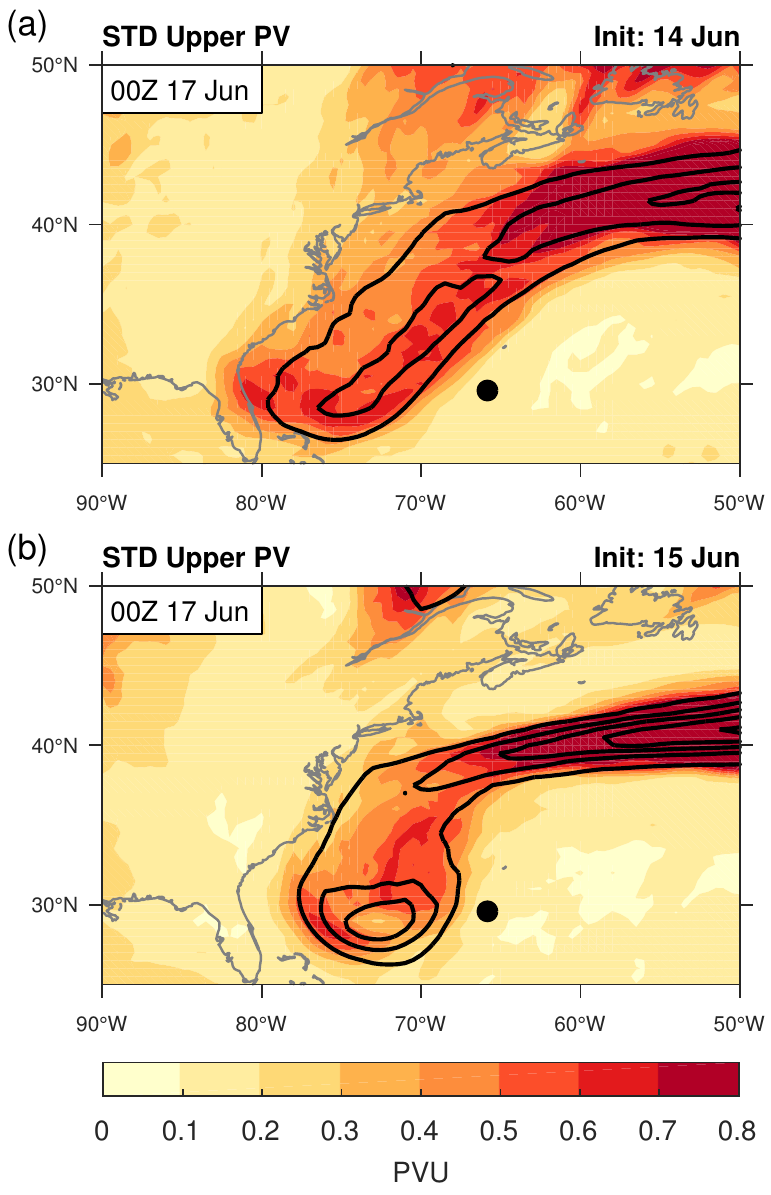}\\
	\caption{Standard deviation of 500--250-hPa averaged PV (shaded, PVU) of the ensemble forecasts from (a) 0000 UTC 14 June 2012 and (b) 0000 UTC 15 June 2012, both valid at 0000 UTC 17 June 2012. The black ensemble-averaged PV contours (at 0.5 PVU intervals starting at 1 PVU) serve as reference and the black-filled circle marks the location of Chris in the analysis.}
	\label{f07}
\end{figure}

Although the predictability of the merging process considerably improves after 12 June, a marked, concomitant rise in the number of similar tracks fails to materialize. Despite a slight increase between 12 and 13 June, the number of similar tracks still does not exceed the 21 members identified in the forecast from 10 June (Fig.\,\ref{f05}). Only with the 15 June initialization, when the merging process was imminent, the ensemble statistics show a prominent rapid change from 16 to 37 members. To elucidate what caused this considerable doubling between 14 and 15 June, ensemble mean and standard deviation of the 500--250-hPa layer-averaged PV is shown for the time when the pre-Chris cyclone develops (0000 UTC 17 June) in Fig.\,\ref{f07}. Comparing the shapes of the ensemble-averaged PV streamers, the forecasts from 14 June are broader and more positively tilted, with an elongated maximum in the middle of the filament (Fig.\,\ref{f07}a). Predictions from the 15 June initialization more closely resemble the PV streamer identified in the analysis, with less implied westerly shear over the developing storm. A "notch" in the PV field northwest of the low center suggests that diabatic PV modification is already under way in many of the members by this time \citep[cf. Fig.\,\ref{f07}b and Fig.\,3 of][]{Davis2004}. Regarding the western end of the PV streamer (west of \SI{65}{\degree}W), the ensemble forecast from 14 June is characterized by high standard deviations along its entire equatorward flank, whereas similar values are found in a substantially smaller area between the PV maximum at the tip and the zonal part of the PV streamer in the forecast from 15 June (Fig.\,\ref{f07}b). In this context, it is remarkable that the highest standard deviations are no longer co-located with the strongest PV gradients but with the area, where diabatic PV redistribution is expected. The spatial confinement of uncertainty to the center of rotation suggests that most of the forecast members agree on the position and overall structure of the hook-shaped PV trough. Thus, subsequent to the merging process, it is the prediction of the actual shape and position of the merged PV streamer that constitutes a dynamical factor limiting the predictability of the pre-Chris cyclone's formation.

\begin{figure}[h!]
	\centering
	\noindent\includegraphics[width = 0.45 \textwidth ]{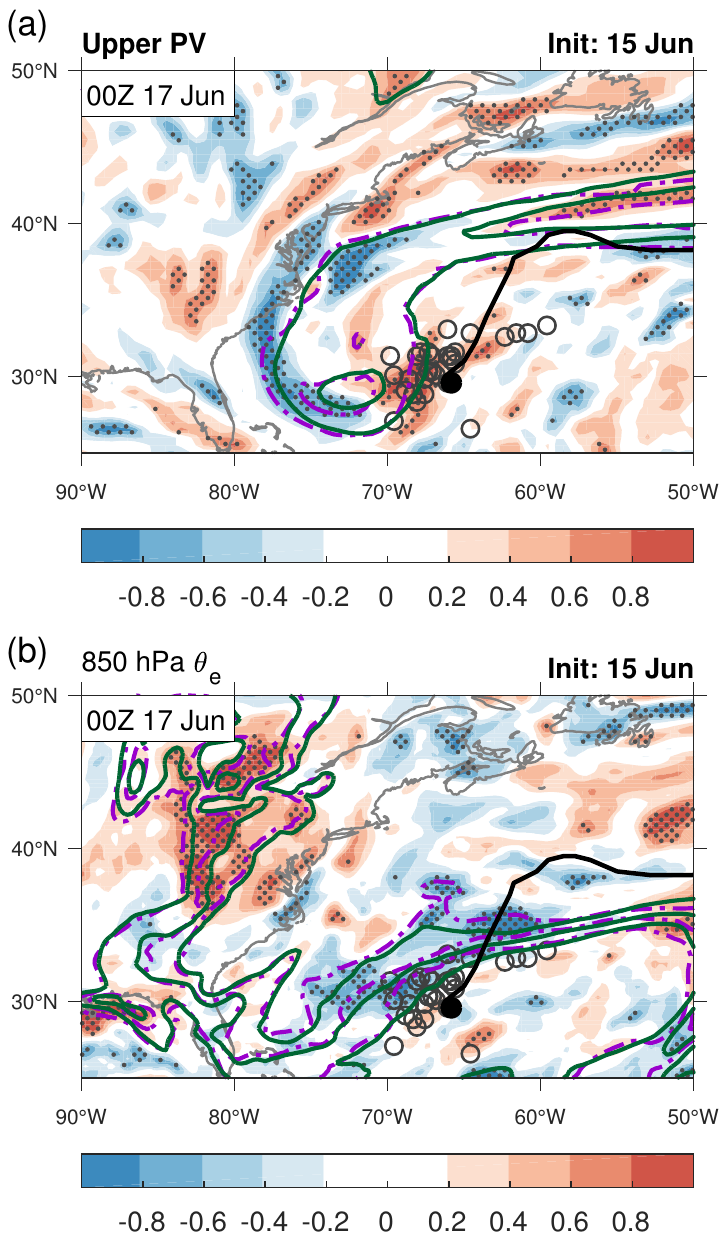}\\
	\caption{(a) as in Fig.\,\ref{f06}, but for the ensemble forecasts from 0000 UTC 15 June 2012 valid at 0000 UTC 17 June 2012. (b) as in (a), but for $\theta_e$ at 850 hPa (\si{\kelvin}) and with the solid green and dotdashed purple $\theta_e$ contours (at 322, 326, and 330 K) indicating where the strongest gradients are located in the \textit{cyclone} and \textit{no-cyclone} groups, respectively. Open (filled) circles mark the cyclone locations in the \textit{cyclone} group forecasts (analysis) at that time and the thick, black line shows the analysis track.}
	\label{f08}
\end{figure}

The ensemble forecast initialized on 15 June is further explored in terms of differences between the upper-level PV streamers in the \textit{cyclone} and \textit{no-cyclone} groups to identify structural characteristics that promoted the development of Chris. Both groups exhibit the hook-shaped PV streamer with a distinct maximum at the southern tip (Figs.\,\ref{f08}a), which suggests that the accurate prediction of the PV streamer's shape is insufficient for forecasting the pre-Chris cyclone's formation. The intensity of the PV maximum is similarly forecasted, but a zonal dipole in the group differences reveals that the PV trough is shifted further east in the \textit{cyclone} group, closer to the formation location of the pre-Chris cyclone. As indicated by the open circles, the majority of the developing storms (except for 5 members) is predicted to emerge between the location of Chris in the analysis and the PV streamer, co-located with significant positive differences. Therefore, it is the combination of the reduced vertical wind shear (cf. Fig.\,\ref{f07}a,b), and the higher upper-tropospheric PV at the formation locations in the \textit{cyclone} group that provides a more favorable environment for the prediction of the pre-Chris cyclone.

Examining the ensemble members from 15 June with regard to low-level thermodynamics, it becomes apparent that the location of the strongest $\theta_e$ gradient at \SI{850}{hPa} is significantly different between the \textit{cyclone} and \textit{no-cyclone} groups (Fig.\,\ref{f08}b). A broad area of significant negative differences appears to the north and west of the developing cyclones, meaning that the location of the north-northwest to south-southeast $\theta_e$ gradient in the \textit{no-cyclone} group is shifted, more distant from the formation locations in the \textit{cyclone} group. The pattern suggests that there is enhanced baroclinicity underneath the eastern side of the upper-level PV streamer in the \textit{cyclone} group compared to the \textit{no-cyclone} group (investigation of corresponding $\theta$ plots confirms the comparison of baroclinicity). The superposition with the higher upper-tropospheric PV (Figs.\,\ref{f08}a) thus provides a thermodynamically more favorable environment for convection.

\subsection{Predictability of the TT of Chris}
\label{ssec:predictTT}

\begin{figure*}[t!]
	\centering
	\noindent\includegraphics[width = 0.9 \textwidth ]{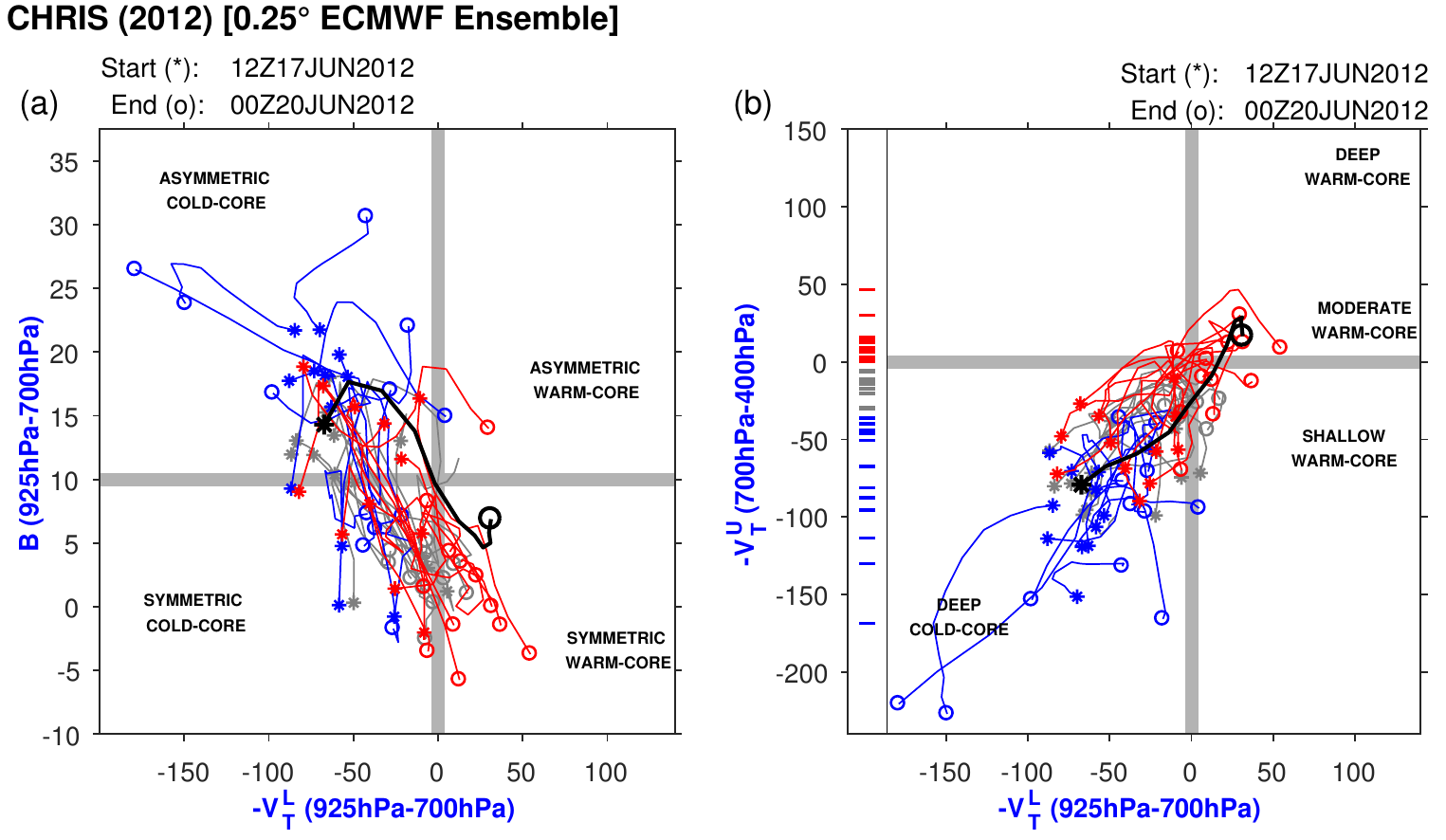}\\
	\caption{(Same as Fig.\,\ref{f04}, but for the \textit{cyclone} group storms in the ensemble forecast from 0000 UTC 15 June 2012 (colored) and for the analysis (black). Trajectories are shown from 1200 UTC 17 June 2012 to 0000 UTC 20 June 2012. The dashes in the left column of (b) show the distribution of the maximum VTU values reached between 0000 UTC 19 June 2012 and 0000 UTC 20 June 2012, separating the \textit{warmer-core} (red), \textit{intermediate-core} (gray), and \textit{colder-core} (blue) terciles (see section \ref{ssec:predictTT} for more details); each trajectory is correspondingly colored. Stars and circles denote the start and end points of the tracks, respectively.}
	\label{f09}
\end{figure*}

For all forecasts initialized later than 14 June, the majority of the ensemble members features a similar track, and thus predicts the development of the pre-Chris cyclone (Fig.\,\ref{f05}). Therefore, the focus shifts to the predictability of the TT, and the 15 June initialization is investigated applying the partitioning strategy described in section \ref{ssec:CPSPartitioning}. Predicted storms in the 37 members of the \textit{cyclone} group split into three terciles (12 \textit{warmer-cores}, 13 \textit{intermediate-cores}, and 12 \textit{colder-cores}) based on the maximum $-V_{T}^U$ values reached between 0000 UTC 19 June and 0000 UTC 20 June. The results described here are qualitatively insensitive to changes in the size of the composite groups. The range of upper level warm core amplitudes is shown along the ordinate in Fig.\,\ref{f09}b. Because the maxima of the \textit{warmer-core} (\textit{colder-core}) tercile are all positive (negative), these two groups of the 15 June initialization can be specified more precisely and are thus referred to as "\textit{transition}" and "\textit{no-transition}" groups hereafter. It is worth noting that the CPS-based partition method captures coherent cyclone-relative differences in storm structure (not shown).

A clear distinction in temporal evolution can be seen between the \textit{transition} and \textit{no-transition} groups in the CPS (Fig.\,\ref{f09}). Compared to the analysis trajectory, most of the trajectories of the \textit{no-transition} storms start as more intense cold cores while the \textit{transition} storms have weaker cold cores on 17 June (Fig.\,\ref{f09}b). Over the subsequent two and a half days, the cores of the circulations in the \textit{transition} group warm throughout the column. Those of the \textit{no-transition} category exhibit mixed behavior that ranges from the rapid warming of initially extreme cold-core circulations, to the deep cooling of more moderate initial structures. Concerning the storm symmetry (Fig.\,\ref{f09}a), most of the trajectories of the \textit{transition} group tend towards decreasing B, reaching values that represent a symmetric structure. In contrast, there is large variability in where the trajectories of the \textit{no-transition} group end. About half of the storms attain a symmetric structure whereas the other half becomes asymmetric.

\subsubsection{Environmental influences}
\begin{figure}[t!]
	\centering
	\noindent\includegraphics[width = 0.42 \textwidth ]{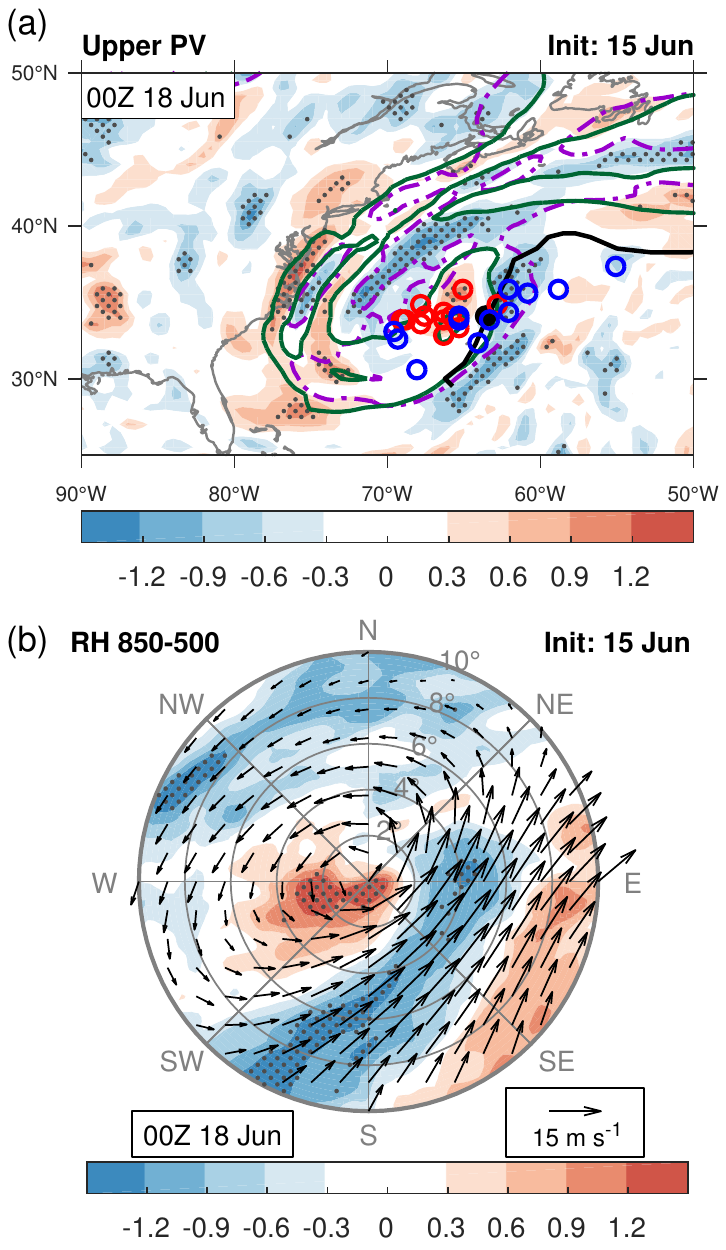}\\
	\caption{(a) Normalized difference of 500--250-hPa layer-averaged PV (shaded, units: standardized anomaly) between the \textit{transition} and \textit{no-transition} groups of the ensemble forecasts from 0000 UTC 15 June 2012 valid at 0000 UTC 18 June 2012. The solid green \textit{transition} group-averaged and dotdashed purple \textit{no-transition} group-averaged PV contours (at 1 and 2 PVU) serve as reference. Red, blue, and black circles mark the cyclone locations in the \textit{transition} group, \textit{no-transition} group, and the analysis, respectively, at that time and the thick, black line shows the analysis track. (b) Normalized difference of cyclone-relative 850--500-hPa layer-averaged relative humidity (shaded, units: standardized anomaly) between the \textit{transition} and \textit{no-transition} groups of the ensemble forecasts from 0000 UTC 15 June 2012 valid at 0000 UTC 18 June 2012. \textit{Cyclone} group-averaged wind vectors (\si{m.s^{-1}}) are calculated for the same layer and scale with the reference vector in the bottom right. Differences significant at the 95\% confidence level are indicated by gray stippling in (a) and (b).}
	\label{f10}
\end{figure}

Different dynamical scenarios are found at 0000 UTC 18 June for the PV streamer associated with the \textit{transition} and the \textit{no-transition} groups, one and a half days before the tropical stage in the analysis (Fig.\,\ref{f10}). The \textit{transition} group features a narrow, wrapping PV streamer, with mainly positive group differences appearing within the 1-PVU contour and significantly reduced PV in the middle of the hook (Fig.\,\ref{f10}a). On the other hand, the PV trough in the \textit{no-transition} group forms a broad, relatively incoherent structure, with significantly lowered differences, equivalent to higher upper-level PV, at its eastern flank. In contrast to the well-defined filament structure in the \textit{transition} composite, the western part of the PV streamer has already degenerated. The \textit{transition} storms predominantly evolve underneath or at the inner side of the narrow PV streamer, which is consistent with a composite study of 2004--2008 North Atlantic TT cases \citep{Galarneau2015}. However, the \textit{no-transition} storms tend to be located at the eastern edge of the broad PV trough, collocated with the area of reduced PV differences (i.e. higher PV) in this group. Another conspicuous difference between the groups is that the positions of the \textit{transition} storms lie close together while those of the \textit{no-transition} storms are spread along the leading edge of the streamer.

\begin{figure*}[t!]
	\centering
	\noindent\includegraphics[width = 0.95 \textwidth ]{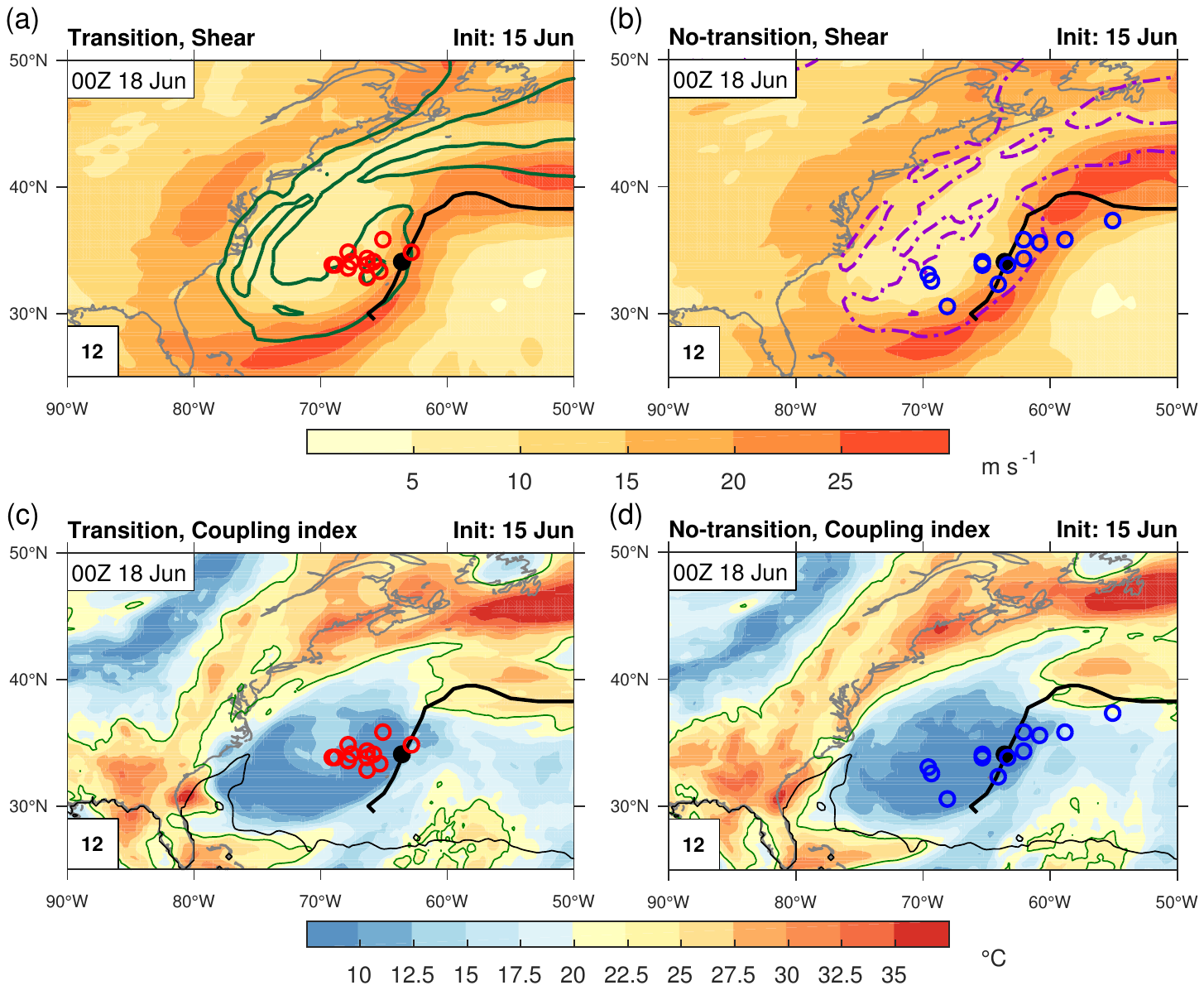}\\
	\caption{Composite mean of (a),(b) 850--300-hPa shear magnitude (shaded, \si{m.s^{-1}}), and (c),(d) coupling index (shaded, \si{\celsius}) for (a),(c) the \textit{transition} and (b),(d) the \textit{no-transition} groups of the ensemble forecast from 0000 UTC 15 June 2012 valid at 0000 UTC 18 June 2012. In (a),(b), the solid green \textit{transition} group-averaged and dotdashed purple \textit{no-transition} group-averaged PV contours (at 1 and 2 PVU) serve as reference. In (c),(d), the green contours highlight the \SI{22.5}{\celsius}-threshold for the coupling index, and black contours represent the canonical \SI{26.5}{\celsius}-threshold for sea surface temperatures. Red, blue, and black circles mark the cyclone locations in the \textit{transition} group, \textit{no-transition} group, and the analysis, respectively, at that time and the thick, black line shows the analysis track. The number at the bottom left corner of each panel denotes the composite size.}
	\label{f11}
\end{figure*}

The predicted positions of the storms relative to the PV streamers determine to what extent they are exposed to the detrimental effect of vertical wind shear. Because the area within the roll-up of the PV streamer is associated with weak 850--300-hPa wind shear (\SIrange{5}{20}{m.s^{-1}}), the \textit{warmer-cores} occur in an environment that is more favorable for the sustained organization of deep convection required for TT (Fig.\,\ref{f11}a). The \textit{colder-cores}, however, experience higher shear along the eastern side of the broad upper-level PV structure, with magnitudes exceeding \SI{20}{m.s^{-1}} (Fig.\,\ref{f11}b).

To examine thermodynamic distinctions in the ensemble forecast from 15 June, we focus on relative humidity because it complements the thermal differences already considered in the $-V_{T}^U$-based separation by a moisture perspective. The most prominent aspect of Figure\,\ref{f10}b is again associated with the area within the wrapping PV streamer: the lower-to-middle troposphere is more moist for the \textit{transitions} compared to the \textit{no-transition} composite. An isolated area of significantly higher relative humidity is found for the \textit{transition} group just west of the center positions (Fig.\,\ref{f10}b). The more pronounced hook-shape of the PV trough in this group helps the moisture to be concentrated in this area. By contrast, the \textit{no-transition} storms are surrounded by a drier environment in the middle of the broad PV structure and more moist conditions (negative differences) along the eastern flank of the PV streamer. Thus, as noted for the TT of Chris in the analysis, the cyclonic roll-up of the PV streamer in the \textit{transition} group causes the seclusion of a warm and moist air mass, leading to deep warming of the core in an environment conducive to TT (cf. Figs.\,\ref{f03}d,f). Less moist conditions result from the predicted PV dynamics in the \textit{no-transition} group, so the storms fail to transition to a TC. Inspection of the composites up to \SI{12}{\hour} before and after 0000 UTC 18 June confirms that all (thermo-)dynamic patterns remain qualitatively steady over this period.

These thermodynamic differences in the local environment are even more remarkable in the light of the group composites for the coupling index \citep{Bosart1995}, which is calculated as the difference between potential temperature ($\theta$) on the dynamic tropopause at 2-PVU and $\theta_e$ at \SI{850}{hPa}. Because its definition links the upper and lower levels, the coupling index assesses the convective stability in the free atmosphere \citep{McTaggart-Cowan2015}. In both subsets, the storms are located in areas, where the coupling index drops to values of less than \SI{10}{\celsius} (Fig.\,\ref{f11}c,d). This is well below the \SI{22.5}{\celsius} maximum threshold proposed by \cite{McTaggart-Cowan2015} as an alternative to the canonical \SI{26.5}{\celsius}-SST-based threshold for TC genesis \citep{Gray1968}. From a (thermo-)dynamic perspective, the crucial point in terms of the predictability of Chris' TT at four and a half days of lead time is therefore not whether the forecasts predict climatologically favorable conditions in terms of convective stability, but rather how the warm and moist air mass is deformed by the PV merger and resulting streamer roll-up process.

\subsubsection{Convective organization}
The distinctly predicted thermodynamic environments, in which the \textit{transition} and \textit{no-transition} storms develop, further provide different conditions for the organization of moist convection that is necessary for a successful TT \citep{Davis2004}. Even though the ECMWF model considered in this study deploys parameterization schemes for convection and boundary layer processes, and absolute values of parameterization-based variables are thus less reliable, differences between the partitioned groups still should be consistent with the thermodynamic scenarios described previously.

\begin{figure*}[h!]
	\centering
	\noindent\includegraphics[width = 0.9 \textwidth ]{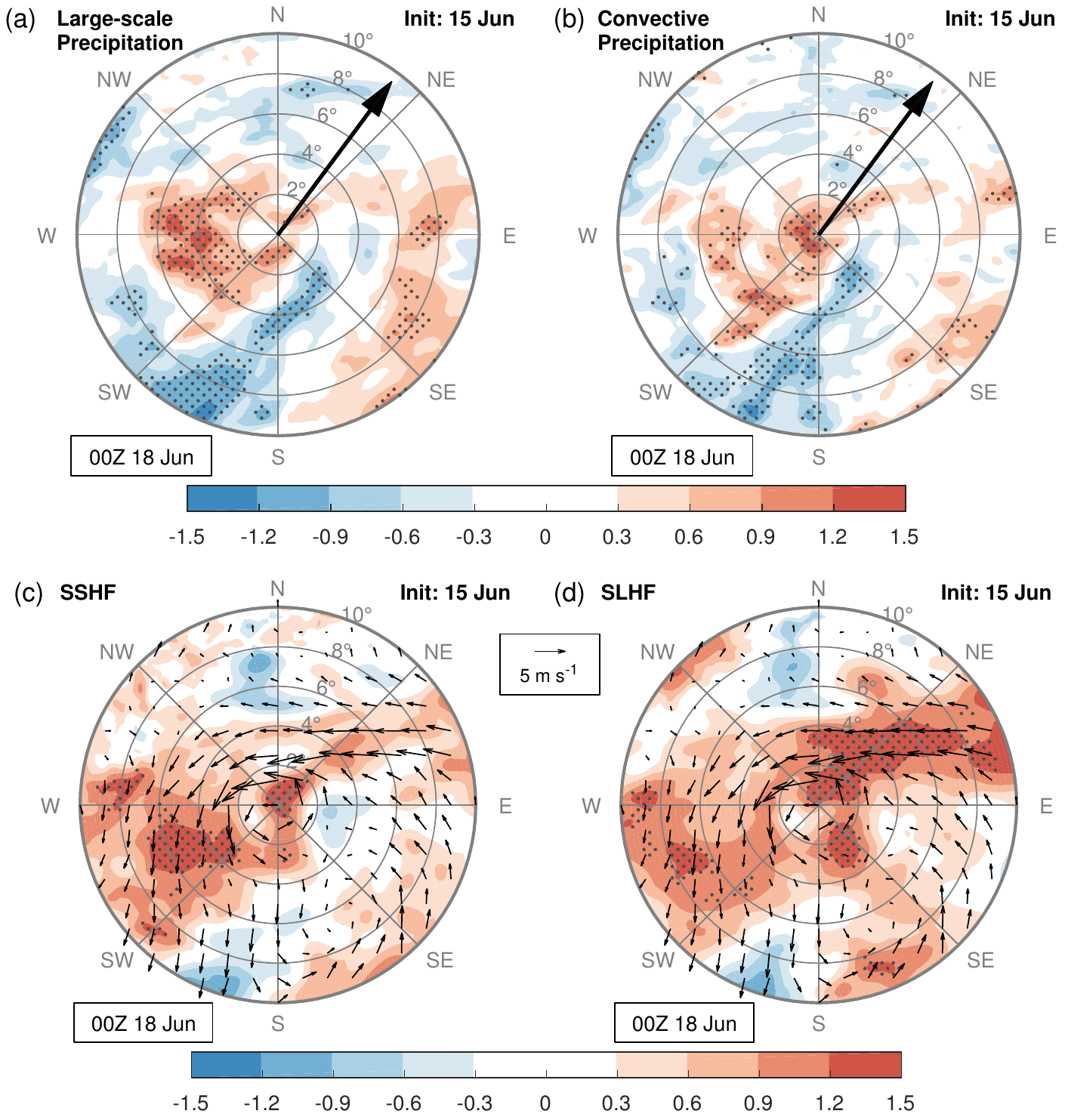}\\
	\caption{As in Fig.\,\ref{f10}b, but for (a) large-scale precipitation rate (\si{\milli\metre\per\hour}), (b) convective precipitation rate (\si{\milli\metre\per\hour}), (c) surface sensible heat fluxes (\si{\watt\per\metre\squared}), and (d) surface latent heat fluxes (\si{\watt\per\metre\squared}). Differences significant at the 95\% confidence level are indicated by gray stippling. \textit{Cyclone} group-averaged shear vectors are represented by thick black arrows in (a) and (b). In (c),(d), vectors show normalized differences between the group-averaged 925-hPa winds, which scale with the reference vector in the middle.}
	\label{f12}
\end{figure*}

In Figure\,\ref{f12}a,b, composites of cyclone-relative differences in precipitation rates in the forecast initialized on 15 June are presented as a proxy for differences in moist convection. As the large-scale precipitation rate suggests, the two distinct PV structures in the upper-troposphere are associated with significant differences in the quasi-balanced forcing for ascent (Fig.\,\ref{f12}a). The warm and moist area enclosed by the PV streamer (cf. Fig.\,\ref{f10}b) features significantly higher large-scale precipitation rates, i.e. stronger upward motions, in the \textit{transition} composite, compared to the \textit{no-transition} group (cf. Fig.\,\ref{f10}b). The overall pattern of differences in the convective precipitation is similarly predicted, however, with stronger signals within a radius of \SI{2}{\degree} (Fig.\,\ref{f12}b). A comparison of the absolute precipitation rates of both variables reveals that the convectively generated vertical motion dominates over the large-scale ascent (not shown). Because of the upshear position of the stronger convection in the \textit{transition} group, the divergent outflow aloft results in a more substantial reshaping of the upper-level PV trough \citep{Davis2003, Davis2004}. As a consequence of this, vertical wind shear is reduced to a greater extent providing a storm environment more favorable for TT.

The combination of surface heat fluxes and low-level wind vectors in Figure\,\ref{f12}c,d indicates the origin of the warm and moist air mass in the lower-to-middle troposphere that becomes secluded by the PV streamer in the \textit{transition} group. The large area of stronger precipitation, and thus enhanced upward motion, west of the \textit{transition} storm centers (Fig.\,\ref{f12}a,b) appears to be in high spatial congruence with significantly increased surface sensible heat fluxes from the ocean into the atmosphere (Fig.\,\ref{f12}c). By contrast, the most striking feature for differences in surface latent heat fluxes is linked to the warm seclusion for the \textit{transition} group. A narrow band of significant positive differences to the northeast is consistent with westward low-level moisture transport by the along-front flow as indicated by the easterly wind vector differences. This resembles the feature found to be associated with transitioning storms in a multi-case study from \cite[][cf. Fig.\,3d,e and 9e]{Galarneau2015}, and suggests that the high $\theta_e$-air in the middle troposphere builds from low-level moisture convergence in the region of stronger upward motion during the seclusion process. It is apparently the lack of surface latent heat fluxes that primarily prevented the \textit{no-transition} storms to acquire organized and sustained convection for an amplification and transition into a TC.

\subsection{Predictability of structural evolution}

\begin{figure*}[h!]
	\centering
	\noindent\includegraphics[width = 0.9 \textwidth ]{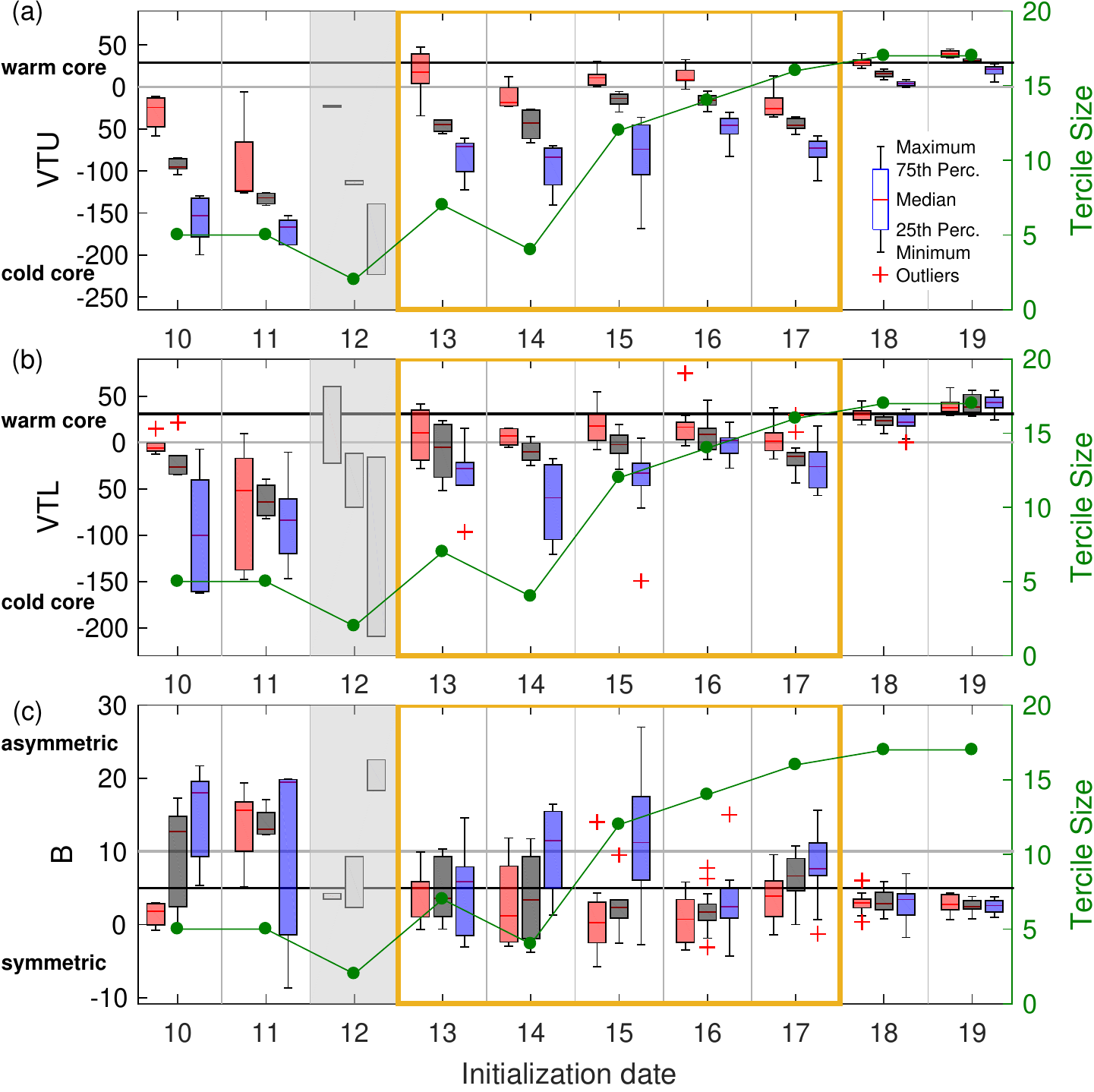}\\
	\caption{Statistical overview of how the ensemble forecasts from 10 to 19 June 2012 predict (a) the maximum $-V_{T}^U$, (b) the maximum $-V_{T}^L$, and (c) the minimum $B$ metric reached between 0000 UTC 19 June and 0000 UTC 20 June. Similar to the approach in Fig.\,\ref{f09}, ensemble members are separated into warm (red), intermediate (gray), and cold (blue) terciles based on the $-V_{T}^U$ metric. Since the terciles from 12 June only consist of two members, the results of this ensemble forecast lack robustness and are hence disregarded (light gray shading). For each category, a box-and-whisker plot displays the median (red line), the interquartile range (IQR, box-length), the most extreme values not considered as outliers (whiskers, maximum 1.5 times the IQR), and the outliers (red crosses). Horizontal black lines indicate the analysis values at 1200 UTC 19 June, and the green lines show how tercile sizes change with lead time. Yellow boxes indicate the most relevant initialization dates.}
	\label{f13}
\end{figure*}

Following the previous examinations of individual verification times for the ensemble forecast initialized at 0000 UTC 15 June, a systematic investigation of the ensemble forecasts initialized between 10 and 19 June with respect to the CPS metrics, as well as environmental and structural storm properties provide a more general perspective on the predictability of Chris' tropical characteristics. As described in section \ref{ssec:CPSPartitioning}, the CPS trajectories for each ensemble forecast are separated into \textit{warmer-core}, \textit{intermediate-core}, and \textit{colder-core} terciles based on the maximum $-V_{T}^U$ values reached between 0000 UTC 19 June and 0000 UTC 20 June. Using these group memberships, the maximum $-V_{T}^L$ and minimum $B$ values are also determined within the 24-hour period. The main features of the obtained statistics are visualized in Figure\,\ref{f13}. Chris' upper-level warm core is greatly underestimated by all members before the initialization on 13 June, while almost every identified storm undergoes TT from 18 June onward (Fig.\ \ref{f13}a). The distinction between \textit{warmer-cores} and \textit{colder-cores} is thus particularly reasonable for the medium range ensemble forecasts initialized between 13 and 17 June (yellow frame) and so the following discussion will focus on this period.

Because the $-V_{T}^U$ and $-V_{T}^L$ metrics are closely related in this case, the \textit{warmer-core} terciles predominately exhibit warm cores and the \textit{colder-core} terciles cold cores in the lower troposphere as well (Fig.\ \ref{f13}b). The \textit{warmer-core} storms also tend to be more symmetric than the \textit{colder-core} storms throughout the ensemble forecasts from 13 to 17 June (Fig.\ \ref{f13}c). A marked reduction in both the differences between the \textit{warmer-core} and \textit{colder-core} structures, and the range of structures within each group, occurs between 15 and 16 June (lead time of 3.5--4.5 days). This abrupt change in predictability is most noticeable for $-V_{T}^U$ and $B$ (Fig.\,\ref{f13} a and c). These prominent changes in the forecast spread are likely due to the remaining uncertainty that is associated with the prediction of the warm and moist seclusion that limit predictability in the 15 June initialization.

\begin{figure}[t!]
	\centering
	\noindent\includegraphics[width = 0.45 \textwidth ]{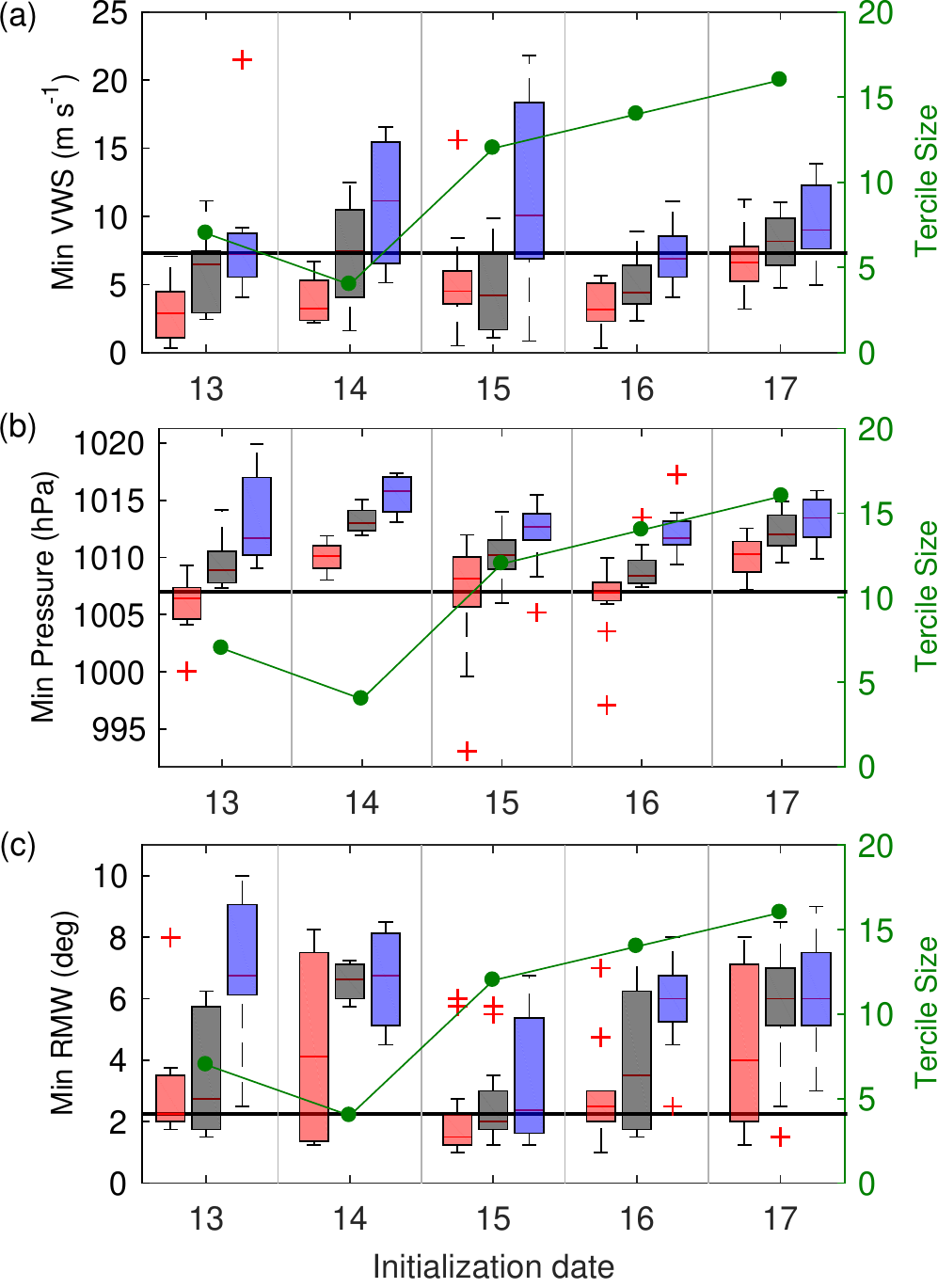}\\
	\caption{Similar to Fig.\,\ref{f13}, but for (a) minimum 850--200-hPa vertical wind shear (\si{m.s^{-1}}) averaged between the radius of \SI{200}{\km} and \SI{800}{\km}, (b) minimum central pressure (hPa), and minimum radius of maximum wind (deg) for ensemble forecasts initialized between 13 and 17 June 2012.}
	\label{f14}
\end{figure}

Applying the $-V_{T}^U$-based group memberships to the environmental shear and to properties indicative of storm structure and intensity, results demonstrate that the \textit{warmer-cores} can be distinguished from the \textit{colder-cores} up to nearly one week before the tropical stage (Fig.\,\ref{f14}). In terms of the 850--200-hPa environmental shear calculated over a \SIrange{200}{800}{\km} annulus, the majority of the \textit{warmer-core} storms experience favorable conditions as shear drops to values generally less than \SI{10}{m.s^{-1}} (Fig.\,\ref{f14}a). For the 17 June initialization, there is an increase in the \textit{warmer-core} group shear caused by the incipient baroclinic interaction of the low-level cyclone with the upper-level PV streamer (cf. B metric in Fig.\,\ref{f04}(a)), consistent with decreased estimates of warm core strength (Fig.\,\ref{f13}a,b). The \textit{colder-core} group consistently predicts a much larger range of shear, in particular for 14 and 15 June, with detrimental shear magnitudes of more than \SI{20}{m.s^{-1}}. Because the vertical axis of a TC gets tilted when strong shear is imposed, these changes in the forecast uncertainty are consistent with the ones described by the $B$ metric (see Fig.\,\ref{f13}c). Using the minimum MSLP for intensity, the \textit{warmer-cores} are readily distinguished from the \textit{colder-cores} because the former acquire significantly lower minima in most of the ensemble forecasts (Fig.\,\ref{f14}b). Only minor differences are apparent in the ensemble statistics for lower-level wind speed between the \textit{warmer-core} and \textit{colder-core} groups (not shown) despite the clear differences observed in Fig.\,\ref{f12}c,d. Instead, the radius of maximum wind describes the compactness of the forecasted storms (Fig.\,\ref{f14}c). Despite the fact that the \textit{warmer-core} storms from 14 and 17 June have neutral or weak warm cores (Fig.\,\ref{f13}a,b), their wind fields are more compact than those of the \textit{colder-core} group, further suggesting that the \textit{warmer-core} storms exhibit a more tropical structure. Similar to the CPS metrics, the spread of the ensemble forecasts collapses markedly after the upper-level PV features merged by early 16 June; however, variance within the groups increases again during the baroclinic interaction of the storm with the upper-level PV streamer.

\section{Discussion and Conclusions}
\label{sec:Discussion}
The aim of the current case study is to systematically investigate a TT event and to identify major limitations for predictability of a) the occurrence of the pre-Chris cyclone, b) its TT, and c) the structural evolution from an ensemble perspective. For this purpose, North Atlantic Hurricane Chris (2012) was chosen because of the complex antecedent PV dynamics and the strong baroclinic environment in the upper and lower levels that facilitated the development of the extratropical precursor cyclone. Before the TC emerged, the predictability at different baroclinic stages is limited by a sequence of events: i) anticyclonic Rossby-wave breaking, ii) the merger of vortex-like PV features, and iii) the cyclonic roll-up of the resultant PV streamer. Ranging from synoptic-scale PV dynamics to differences in the convective organization, this study seeks to provide a better understanding of the main sources of uncertainty that are associated with these atmospheric features and processes across a broad range of scales.

The results of this investigation shows that the predictability of the pre-Chris cyclone's occurrence is strongly related to the predictability of the preceding PV dynamics. At five-to-six-day lead times (11 and 12 June) prior to the development of the non-tropical precursor cyclone, formation of the pre-Chris cyclone was only predicted by ensemble members that successfully merged pre-existing PV remnants and a cut-off low to develop the precursor PV trough. Once the majority of the members predicted the PV merger from four days before the pre-Chris cyclone formation on (the 13 June initialization and beyond), the number of similar storm tracks doubled (Fig.\,\ref{f05}). This dramatic increase in predictability appears to be related to the fact that the anticyclonic wave break has already generated the critical upper-level cut-off low by this time, and uncertainty associated with the existence of the cut-off low is therefore dramatically reduced. The largest increase in the predictability of storm occurrence, however, takes place between three and two days prior to the occurrence of the pre-Chris cyclone (14 to 15 June), when the bulk uncertainty in the area of the PV trough becomes restricted to the interior of the cyclonic PV streamer roll-up. This regime change in predictability is found to be attributable to the PV merging that was imminent on that day. A more detailed analysis of the resultant trough structures reveals that the members predicting the occurrence of the pre-Chris cyclone are linked to a superposition of higher PV values at upper levels and stronger thermal gradients at lower levels, i.e. an overall stronger baroclinic environment. These findings corroborate the conjecture of \cite{Wang2018} that the set of factors relevant for tropical cyclogenesis -- e.g. absolute vorticity, relative humidity, potential intensity, and vertical wind shear in the genesis potential index of \cite{Emanuel2004} -- needs to be extended for the TT pathways. Because these factors generally represent tropical ingredients and thus predominantly non-baroclinic conditions, further research is required to determine how the baroclinic precursor dynamics involved in the TT pathways could be incorporated into a conceptual or practical model of TT likelihood.

\begin{figure*}[h!]
	\centering
	\noindent\includegraphics[width = 0.8 \textwidth ]{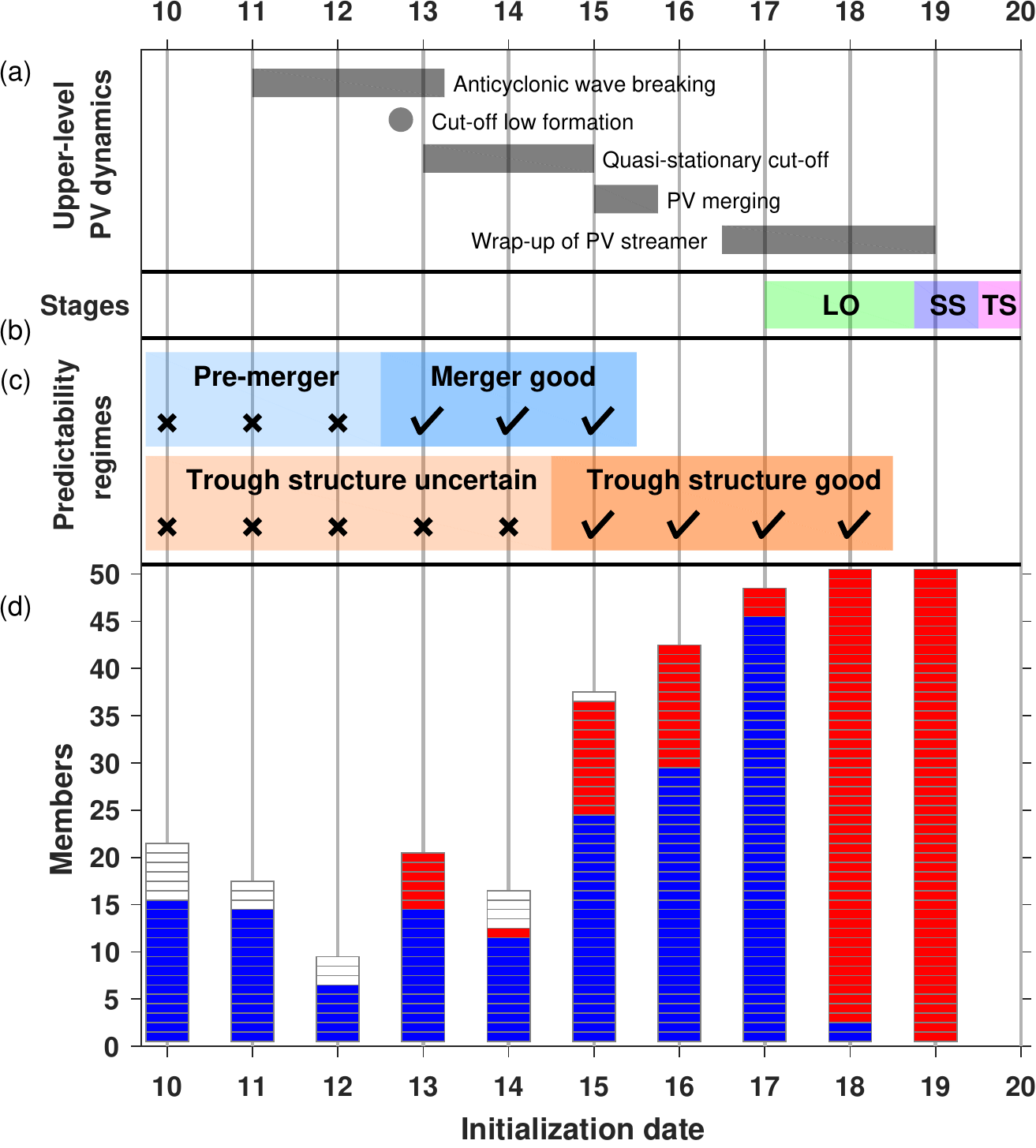}\\
	\caption{Schematic diagram summarizing the predictability of Hurricane Chris between 10 and 19 June 2012: (a) crucial periods (bars) and events (points) in the upper-level PV dynamics of the analysis (cf. section \ref{sec:Overview}), (b) stages in the NHC Best Track, and (c) identified predictability regimes with an assessment of whether the atmospheric process or feature is not (crosses) or is "sufficiently" well (ticks) predicted in the ensemble forecasts. (d) Overview of the ensemble predictive skill, with the height of the stacked bars indicating the number of identified similar tracks and the red (warm core), and blue (cold core) segment colors corresponding to the maximum $-V_{T}^U$ value reached between 0000 UTC 19 June 2012 and 0000 UTC 20 June 2012. Uncolored segments indicate that the similar storm track ends before 0000 UTC 19 June.}
	\label{f15}
\end{figure*}

Because this case study aims to document the predictability of TT, further examination is performed to elucidate why the forecasted developing storms in the ensemble successfully complete TT. Simulations begin to accurately predict TT almost one week before TT occurs, at the time when most ensemble members agree on the PV merger (Fig.\,\ref{f15}). The increased predictability of the PV merger, and hence the formation of the upper-level precursor PV trough, appears to lead to these first TT predictions. It is somewhat surprising that no considerable increase in the proportion of the warm cores is found in the subsequent initializations, until the storm itself has developed (at some lead times, even fewer TTs occur). The majority of ensemble members predict a warm core only after the pre-Chris cyclone became located underneath the PV streamer one and a half days before the tropical phase. This study confirms the findings of \cite{Majumdar2014}, who also identify reduced predictive skill in ECMWF short-range ensemble predictions of 2010--2012 North Atlantic warm core formations.

Composite differences between the warmest and coldest upper-level cores in the forecast initialization from 0000 UTC 15 June are analyzed during the roll-up of the PV streamer to highlight the (thermo-)dynamics that assist or imped the storm's completion of TT. From a dynamical point of view, the shape of the PV streamer in combination with the relative storm position are most decisive in determining whether TT occurs. Similar to the findings of \cite{Galarneau2015}, the \textit{transition} storms are located inside a narrow, wrapping PV streamer and interact with the high-PV feature at a shorter distance compared to the \textit{no-transition} storms. The latter are steered northeastward along the leading edge of a broad but incoherent PV structure. Reduced vertical shear characterizes the near-storm environment in the \textit{transition} cases. The distinct upper-level PV dynamics are also related to different thermodynamic scenarios. The narrow PV streamer in the \textit{transition} cases lead to the isolation of a warm, moist air mass primarily in the mid-troposphere, which provide a more tropical environment. Conversely, detrimental dry air ahead of the PV trough is found in the members whose predicted storms fail to complete TT. In terms of convective organization, stronger low-level winds northeast of the \textit{transition} centers induce enhanced surface latent heat fluxes, and thereby transport more moisture into the region of upward motion, confirming the multi-case study results from \cite{Galarneau2015}. In accordance with the modelling study from \cite{Davis2003}, the upstream position of the enhanced convection and the attendant diabatic outflow leads to a stronger reduction in vertical wind shear.

The present study has also shown that it is possible to make a skillful distinction between \textit{warmer-core} and \textit{colder-core} storms in CPS metrics, and environmental and structural storm properties in medium-range forecasts. In agreement with the findings of \cite{Davis2003}, deep layer wind shear below \SI{10}{m.s^{-1}} appears to be favorable for TT. With regard to the CPS metrics, substantial forecast improvements are linked to the end of the upper-level PV merger as well as to the time when the upper-tropospheric PV trough connects to the non-tropical precursor cyclone. This suggests that interactions of baroclinic features prior to TT are major sources of forecast uncertainties for this TC development pathway.

The present study is the first to investigate changes in predictability of a TT event with lead time in order to identify the major limiting factors in the antecedent dynamics. Various atmospheric features and processes are found to significantly affect predictability. The consistency of these case study results with previous climatological investigations suggests that the conclusions drawn here may be more generally applicable, at least in a qualitative sense. Moreover, the methods developed here open a promising avenue to multi-scale predictability studies of TTs in all basins. More research is required to understand better the relative importance of the individual features to predictability, ideally using a feature-based framework. The results of such an investigation will be reported in a future study that will further quantify the predictability of TC formation via the TT pathway.\\

\noindent\textbf{Acknowledgment}\\
\footnotesize
The research leading to these results has been accomplished within project C3 ‘‘Multi-scale dynamics and predictability of Atlantic Subtropical Cyclones and Medicanes’’ of the Transregional Collaborative Research Center SFB/TRR 165 ‘‘Waves to Weather’’ funded by the German Science Foundation (DFG). We would like to thank the editor, two anonymous reviewers, and Alex Kowaleski for their critical and constructive comments which helped to improve significantly the quality of the paper. The authors also thank Andreas Schlueter and various other colleagues for helpful discussions.

\bibliography{references}

\end{document}


\begin{center}
			{\bf \Large Supplement to "Tropical transition of Hurricane}\\
			{\bf \Large Chris (2012) over the North Atlantic Ocean:}\\
			{\bf \Large A multi-scale investigation of predictability"}
			
			\bigskip
			\medskip
			
			{\bf Michael Maier-Gerber\textsuperscript{*1}, Michael Riemer\textsuperscript{2}, Andreas H.~Fink\textsuperscript{1},\\ Peter Knippertz\textsuperscript{1}, Enrico Di Muzio\textsuperscript{1,2}, Ron McTaggart-Cowan\textsuperscript{3}}\\

			\bigskip
			{\tiny\textsuperscript{1} Institute of Meteorology and Climate Research, Karlsruhe Institute of Technology, Karlsruhe, Germany}\\
			{\tiny\textsuperscript{2} Institute for Atmospheric Physics, Johannes Gutenberg University, Mainz, Germany}\\
			{\tiny\textsuperscript{3} Numerical Weather Prediction Research Section, Environment and Climate Change Canada, Dorval, Quebec, Canada}
			
			\bigskip
			\footnotesize\textsuperscript{*}michael.maier-gerber@kit.edu
			
			\bigskip
			
			This document is a supplement to "Tropical Transition of Hurricane Chris (2012) over the North Atlantic Ocean: A Multi-Scale Investigation of Predictability". It contains a plot, which served as basis for the choice of a threshold applied to the dynamic time warping technique. This threshold represents the maximum allowed average spatio-temporal discrepancy between forecast and analysis tracks considered as similar. In addition, GOES-13 satellite images are provided for the evolution of Chris prior to tropical transition.

		\end{center}
		\bigskip
		\bigskip
		\bigskip
		\bigskip

\renewcommand{\thefigure}{S\arabic{figure}}
\setcounter{figure}{0} 

\begin{figure*}[h]
	\centering
	\noindent\includegraphics[width = 0.9 \textwidth ]{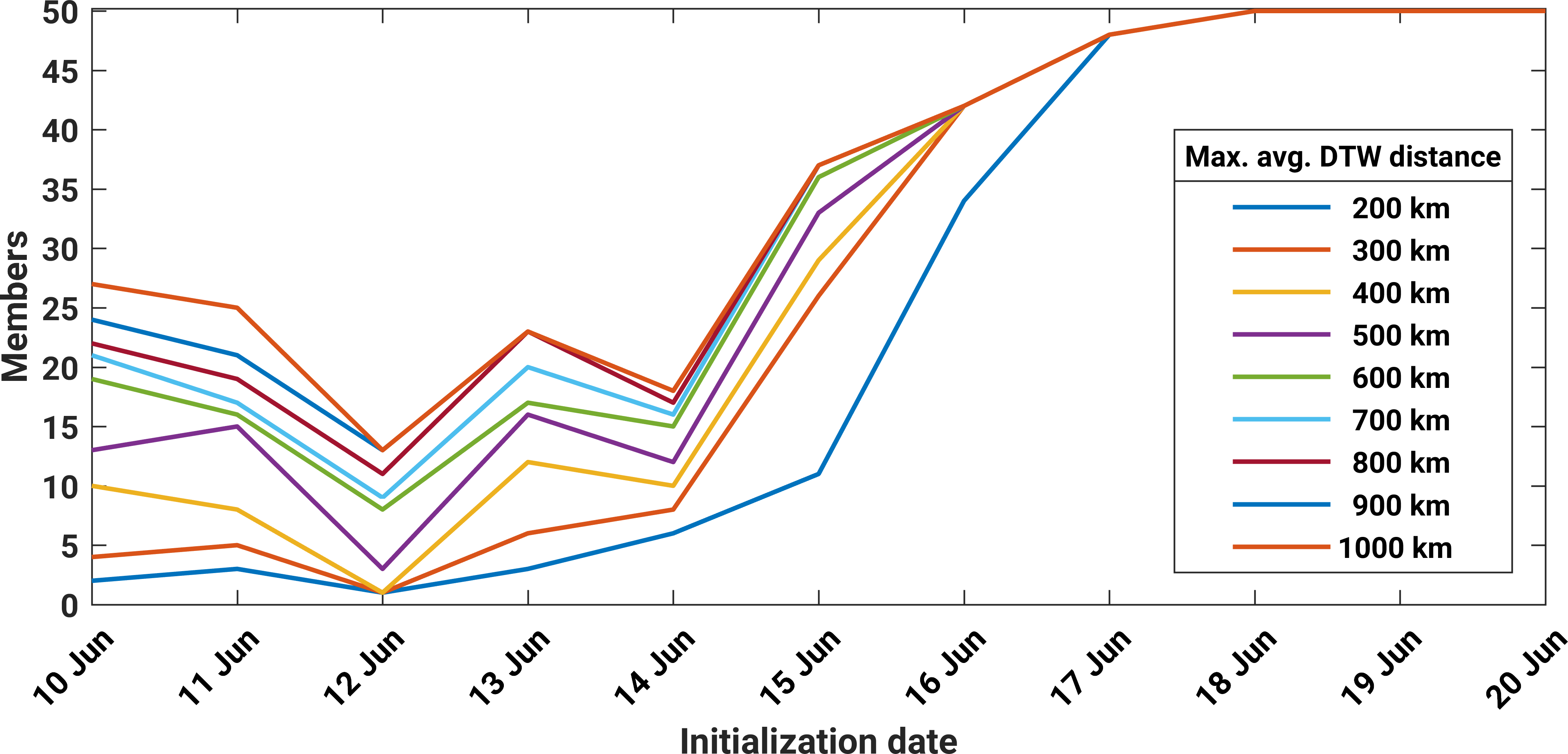}\\
	\caption{Number of ECMWF ensemble members that show a storm similar to Chris for different thresholds (\SIrange{200}{1000}{\km}) of the maximum allowed average spatio-temporal discrepancy between forecast and analysis tracks ($\overline{d_{DTW}}$). All forecasts are initialized at 0000 UTC.}
\end{figure*}

\begin{figure*}
	\centering
	\noindent\includegraphics[width = 0.65 \textwidth ]{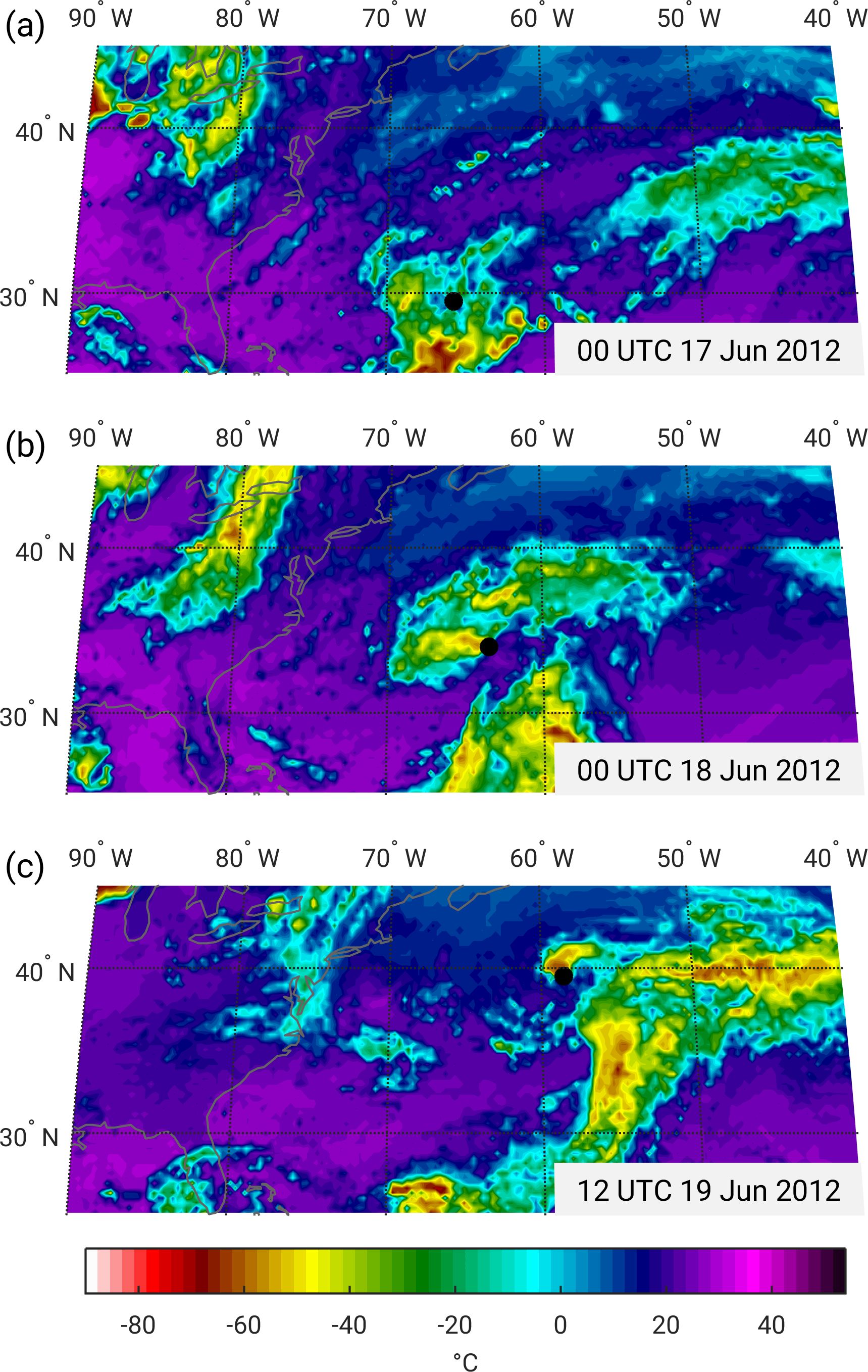}\\
	\caption{GOES-13 thermal infrared satellite images valid at (a) 0000 UTC 17 June 2012, (b) 0000 UTC 18 June 2012, (c) 1200 UTC 19 June 2012, showing the development of Hurricane Chris during the pre-tropical phase. The black dot indicates the storm center position at surface.}
	\label{sf02}
\end{figure*}

\begin{figure*}
	\centering
	\noindent\includegraphics[width = 0.65 \textwidth ]{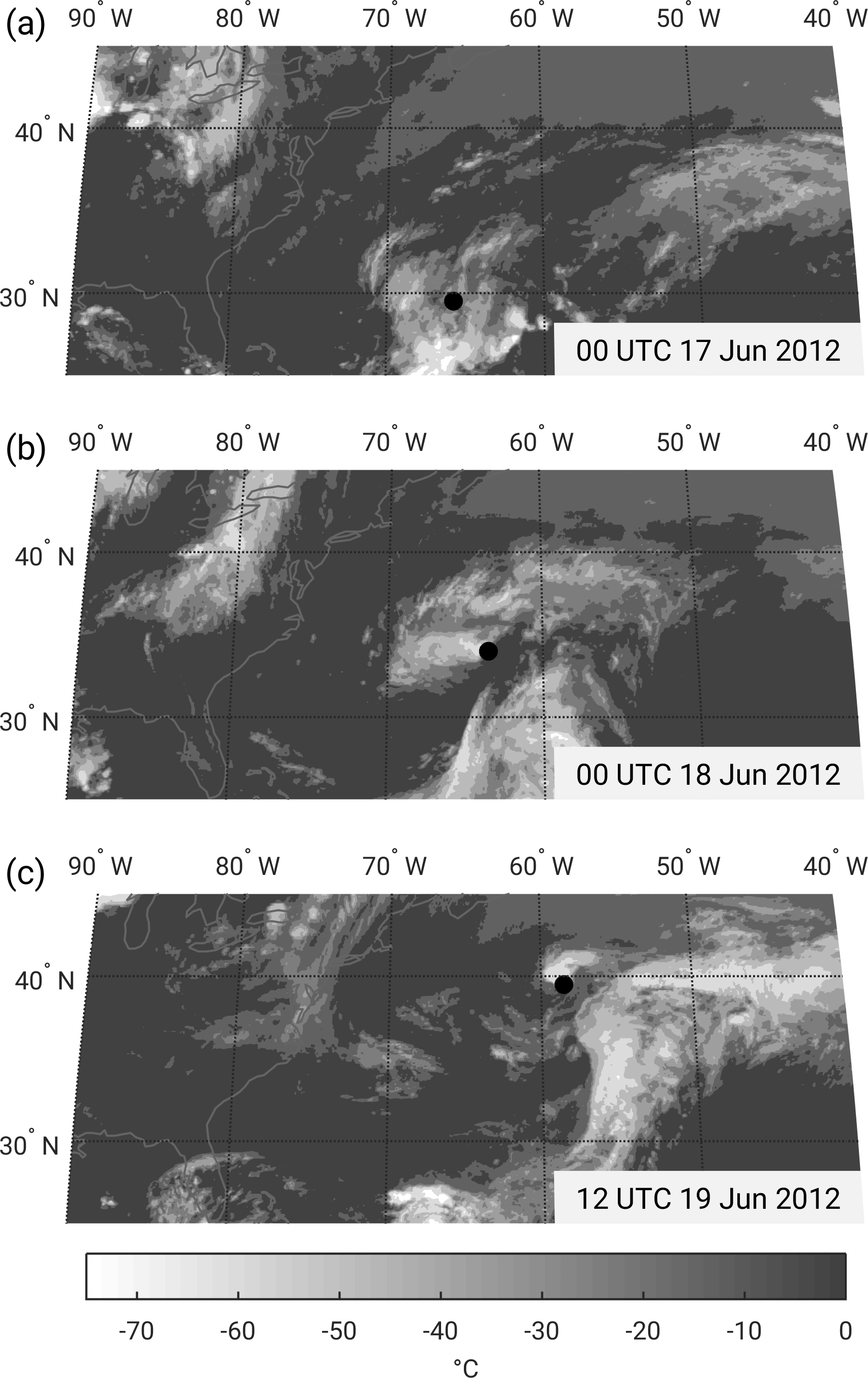}\\
	\caption{Same as Fig.\,\ref{sf02}, but for GOES-13 water vapour satellite images.}
\end{figure*}